\documentclass[twocolumn,tighten]{aastex62}

\hypersetup{linkcolor=red,citecolor=blue,filecolor=cyan,urlcolor=magenta,breaklinks=true}
\usepackage{breakurl}
\usepackage{breakcites}
\usepackage{amsmath}

\shorttitle{Formation of Warm Sub-Neptunes}

\shortauthors{Bodenheimer et al.}

\received{June, 28}
\revised{October, 8}
\accepted{October, 15; To appear in The Astrophysical Journal}
\reportnum{LA-UR-18-23470}

\begin{document}


\title{\uppercase{New Formation Models for the Kepler-36 System}}            


\author[0000-0001-6093-3097]{Peter Bodenheimer}
\affiliation{UCO/Lick Observatory, Department of Astronomy and Astrophysics, University of California,
    Santa Cruz, CA 95064, USA}
\email{peter@ucolick.org}

\author[0000-0001-9432-7159]{David J. Stevenson}
\affiliation{Division of Geological and Planetary Sciences, Caltech, Pasadena, CA 91125, USA}
\email{djs@gps.caltech.edu}

\author{Jack J. Lissauer}
\affiliation{Space Science and Astrobiology Division, NASA-Ames Research Center, 
Moffett Field, CA 94035, USA}
\email{Jack.J.Lissauer@nasa.gov}

\author[0000-0002-2064-0801]{Gennaro D'Angelo}
\affiliation{Theoretical Division, Los Alamos National Laboratory, Los Alamos, NM 87545, USA}
\email{gennaro@lanl.gov}

\begin{abstract}
Formation of  the planets in the Kepler-36 system
is  modeled   by detailed numerical simulations
according to the core-nucleated accretion scenario. The standard model 
is updated to include the dissolution of accreting rocky planetesimals in
the gaseous envelope of the planet, leading to substantial enrichment of 
the envelope mass in heavy elements and a non-uniform  
composition with depth. For Kepler-36 c, models involving in situ formation and models involving
orbital migration are considered.  The results are compared with standard formation
models. The calculations include the formation (accretion) phase, as well as the
subsequent cooling phase, up to the age of Kepler-36 (7 Gyr).  During the latter phase,
mass loss induced by stellar XUV radiation is included. In all cases, the results fit the
measured mass, 7.84 M$_\oplus$,  and radius, 3.68 R$_\oplus$,  of Kepler-36 c. Two 
parameters are varied to obtain these fits: the disk solid surface density at the formation 
location, and the ``efficiency" factor in the XUV mass loss rate.
The updated models are hotter and therefore less dense in the silicate portion of the planet and in the 
overlying layers of H/He, as compared with standard models. The lower densities 
 mean that only about half as much H/He is needed to be accreted to fit the present-day mass and 
radius constraints.  For Kepler-36 b, 
an updated in situ calculation shows that the entire H/He envelope is lost, early in the cooling phase, in
agreement with observation.
\end{abstract}

\keywords{planets and satellites: formation --- planets and satellites:
physical evolution --- planets and satellites: individual (Kepler-36 c, Kepler-36 b)}

\section{Introduction}
\label{sect:intro}

Thousands of  extrasolar planets have been discovered during the past decade.
 A substantial fraction of these were found    through
transit observations by the main 
{\it Kepler} mission, which  identified   more than 4000 
planetary candidates, the majority of which have been verified as true 
exoplanets (\url{https://www.nasa.gov/kepler/discoveries}).
The general observed properties of extrasolar planets are reviewed by \citet{win15} 
and \citet{lis14}. 
Most of the {\it Kepler}  planets orbit within 0.5 AU of their star, and have 
radii  between those of  Earth and Neptune.
A subset of the {\it Kepler} planets also
have mass determinations, as found either by radial velocity measurements of the
transiting planets \citep{mar14}, or by transit timing measurements in systems with multiple planets \citep{ago18}. 
A diagram of the radii and masses of such objects, with radii $R < 4.2~$R$_\oplus$, can be 
found in \citet{kal17}. 
For  example, transit timing variations in the
Kepler-11 system yield masses between 1.9 and 8.0 M$_\oplus$
for planets with  radii between 1.8 and 4.2 R$_\oplus$  \citep{lis13}.

We consider planets in the range 1--10 M$_\oplus$ and radii $R < 6$ R$_\oplus$. 
The  mass and radius measurements give the planetary mean density $\bar \rho$. 
Those with $\bar \rho >5.0$ 
g~cm$^{-3}$ $(M_p/{\rm M}_\oplus)^{0.7}$ 
 must  be composed  almost entirely of heavy elements (primarily rock) with hardly
any hydrogen/helium (H/He) atmosphere. This conclusion is true for planets of 
$R=1$ R$_\oplus$, but larger planets must be more dense for heavy elements to
dominate by volume. 
 The  low-density planets  ($\bar \rho <1.5$ g cm$^{-3}$)   can still have most         
of their mass in a heavy-element core of rock and (possibly) ice, but  they
must also
have a volumetrically significant  outer envelope occupied by light gasses (H and/or He). 
Intermediate-density planets can resemble the low-density planets but with
the outer envelope occupying a smaller fraction of the volume, 
 or they could  be composed mostly  of water and/or other
astrophysical ices. Observationally there appears to be a boundary in radius between those
planets that  are composed (almost) entirely of heavy elements and those with
a light-element envelope. Based on a limited sample of transiting planets with
radial-velocity mass determinations, \citet{rog15} finds that few planets larger than 
1.6~R$_\oplus$ are composed entirely of rock (silicates plus iron). 
Above 2~R$_\oplus$, the planet is very
likely to have a substantial fraction of its volume occupied by light elements.
Further observations indicate a bimodal distribution of planetary radii \citep{ful18,van18}, 
with a definite dip in  the number of planets with radii around 1.8 R$_\oplus$.  Most planets with orbital periods
less than 100 days have radii either $< 1.6$ R$_\oplus$ or 
2--3 R$_\oplus$.

\begin{figure*}[t]
\centering%
\resizebox{0.6\linewidth}{!}{\includegraphics[clip]{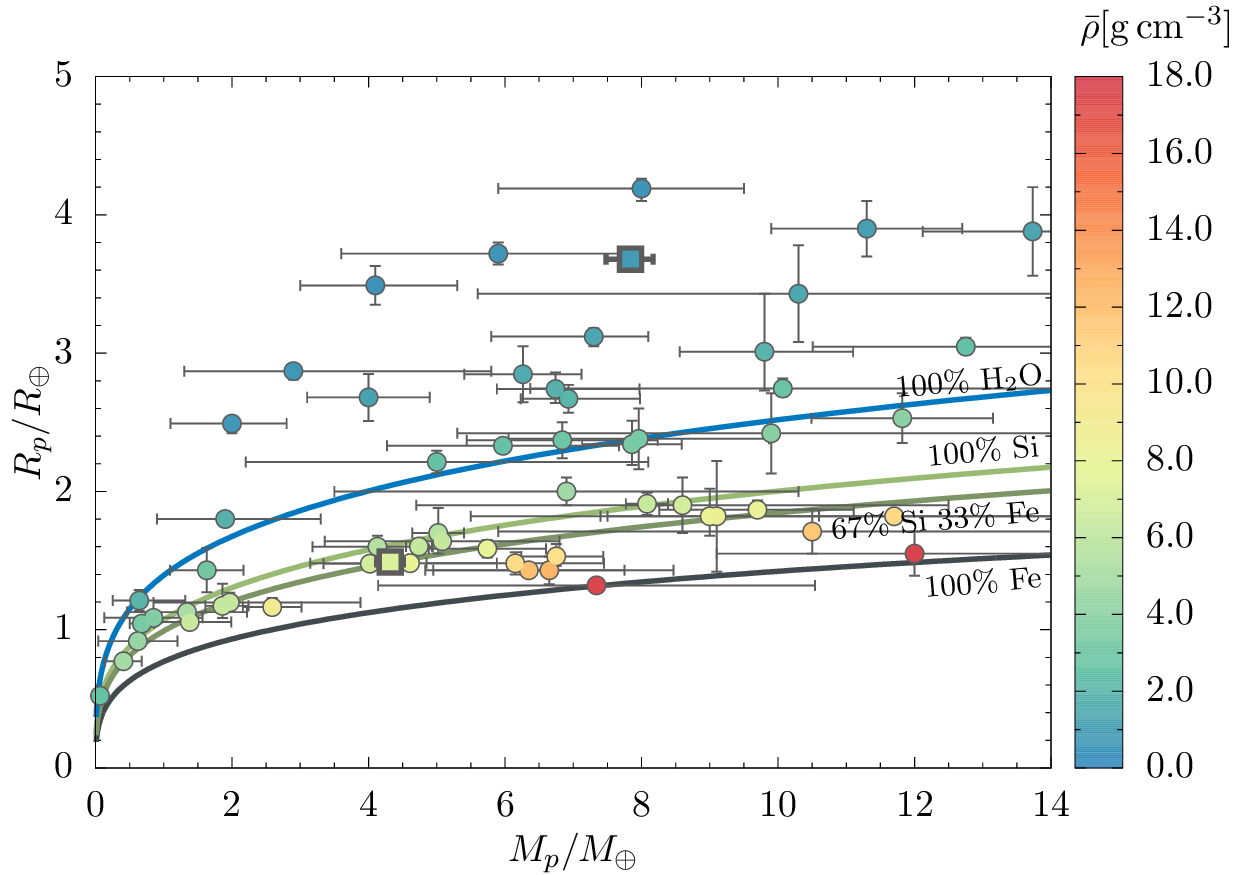}}
\caption{Radius ($R_p$) versus mass ($M_p$) of selected  extrasolar planets whose mass and radius
have both been measured (source: NASA Exoplanet Archive).
Some planets with large error bars have not been plotted. 
The color scale renders the average density,
computed from the mass and radius values. The vast majority of the planets were observed by
the {\it Kepler} mission. The four   curves indicate the radius of solid cores of different
compositions, as indicated \citep{dan16}.  The two
squares    indicate Kepler-36 b (less massive) and c. }
\label{fig:11}
\end{figure*}

A particularly interesting system in this regard is that of the star
Kepler-36 \citep{car12}, an evolved subgiant  with mass 1.07 M$_\odot$ and radius
1.626 R$_\odot$. The planet Kepler-36 b has an orbital period of
13.84 days, a mass of 4.32 (+0.19, $-0.20$) M$_\oplus$, and a radius of 
1.49 $\pm$ 0.035 R$_\oplus$, while its neighbor Kepler-36 c has a period of 16.238 days,  
 a mass of 7.84 (+0.33, $-0.36$) M$_\oplus$, and a radius of 3.68 (+0.056, $-0.055$)
R$_\oplus$ \citep{dec12}. The masses are determined from transit timing variations and refined
by considerations of long-term orbital stability of the system \citep{dec12}. 
The precision of the masses and radii of the two planets is among the best
available for extrasolar planets; thus, this system is a prime target for theoretical
analysis.
Standard structure models \citep{lop13} indicate that planet c is likely to
have an H/He envelope containing about 9\% of the total mass, while planet b
is likely to be a rocky planet with  no H/He envelope, or at most one with less than 0.1\% of the mass.
The mean densities of planets b and c are, respectively, $\approx 7.23 \pm 0.61$ and
$\approx 0.87 \pm 0.055$ g cm$^{-3}$. Mean densities of a number of well-observed planets,
including those of Kepler-36, are shown in Figure \ref{fig:11}. 

Model calculations of the evolution of the Kepler-36  planets,
starting after formation at an age of 10 Myr and ending at the present age
of the star (6.92 $\pm$ 0.37 Gyr) are reported by
\citet{lop13}.  The model planets are located on their current orbits.
The models consist of a heavy-element core, with constant  mass $M_\mathrm{core}$,
equal to the present deduced core masses of the planets, and a H/He
envelope, which cools and loses mass with time as a result of XUV irradiation from
the star. Both planets are assumed to start with a H/He mass fraction of
22\%, and the results show agreement with the current masses and radii of
the planets. Planet b loses its entire H/He envelope, while planet c is
left with an envelope with mass fraction about 9\%. The enhanced mass
loss in planet b is not primarily a result of the slightly higher XUV 
flux (the orbital radii of planets b and c differ by $\approx$ 10\%),
but rather because of the significantly lower $M_\mathrm{core}$, which makes the planet
more susceptible to mass loss.  These authors show that the mass loss
time scale goes roughly as $M_{\mathrm{core}}^2$.

Similar 
calculations were performed by \citet{owe16a}, again 
starting after formation and with the orbits at their present positions.
Some differences in assumptions were made regarding 
the XUV mass loss rate and the dependence of the XUV flux on time.
The conclusion was that planet b started with an envelope mass fraction
of less than about 10\% and that the planet lost its entire envelope,
while planet c started with an envelope mass fraction of 15--30\% and
retained an H/He envelope with mass fraction about 10\%.

This paper investigates the origin and
evolution of the Kepler-36 planets, assuming that they form somewhere in the
inner disk, inside the snow line, according to the core-nucleated accretion
process \citep{saf69,pol96,dan18}.  
In our   past work
\citep{pol96,mov10,rog11,dan16} and references therein,
 the accreting planetesimals orbit through the
gaseous envelope, ablating and breaking up during the process. The
amount of solid material that is deposited at each layer of the
envelope is determined. However, in practically all calculations,
 the heavy-element material is assumed to sink to
the core, leaving the envelope with a composition of pure H/He. A small amount
of dust remains in the atmosphere and is a source of opacity to outgoing
thermal radiation, but only the overwhelmingly dominant elements H and He
are included in the calculations of the equation of state within the envelope.
The main improvement in the present  work involves the fate of the accreted planetesimals, 
which  are allowed to break up, vaporize, and dissolve in the H/He envelope, thereby enriching 
the envelope, non-uniformly, in heavy elements. 

Several previous calculations have considered this effect. 
\citet{ven16} assume that a  planet forms beyond the ice line and accretes planetesimals
composed of rock and ice. All of the rock sinks to the core, and
a fraction of the ice (nominally 50\%) remains in the envelope
while the remainder sinks. The ices in the envelope are uniformly
mixed throughout its entire mass,  so the envelope composition is
uniform at all times, but enriched in ices compared with solar
composition.  The authors find that various types of planets, including 
giant  planets as well as Neptune-type
planets, can be formed through such envelope enrichment. Also, the formation
of gas giants is accelerated by this process, and  the planetary metallicity
is predicted to decrease with increasing planetary mass.

  Further calculations are reported by
\citet{ven17}; again all of the rock component of the planetesimals
sinks to the core, while all of the ice remains uniformly mixed
in the envelope. Formation locations range from 5 to 30 AU, with subsequent
orbital migration, 
and planetesimal accretion as well as pebble accretion is considered.
The object is to determine the occurrence rate of mini-Neptunes,
that is, planets with mass $M_{\rm p} < 10$ M$_\oplus$ and with H/He
mass fractions between 0.1 and 0.25. This occurrence rate is found
to depend on solid particle size, formation location, and envelope opacity.
For low opacity with  pebbles, the rate is found to increase when envelope enrichment is included and 
the formation location is around 20--30 AU. For low opacity with   planetesimals, the same is true if
the formation location is around 5 AU. For high opacity, for both pebbles and planetesimals, the
favored location is at 20--30 AU, and the rate increases  significantly with envelope enrichment.

Formation calculations for Jupiter \citep{loz17} include envelope
enrichment in heavy elements but are based on pre-computed
structure models for the formation of the planet.  It is not assumed that
the rock portion of the rock/ice planetesimals
falls to the core
or that uniform mixing necessarily occurs. Early on, planetesimals
do accrete to form a solid/liquid core, but once this core
reaches 1--2 M$_\oplus$, they dissolve in the envelope, forming a
non-uniform composition distribution. Silicate vapor
tends to concentrate toward the center, and its presence leads to
much higher temperatures in the envelope than those in envelopes
composed of pure hydrogen/helium. Most of the accreted heavy-element
material remains in the envelope and does not settle to the core.
A further calculation for Jupiter, again without the assumption of uniform
mixing, investigates the structure allowing for a gradient in
the mass fraction of heavy elements \citep{hel17}. That gradient
could be quite steep, leading to a fairly well-defined core/envelope
structure, or it could be quite gradual, depending on the history
of the gas accretion rate compared with the solid accretion rate.

\citet{cha17} considers a planet forming at
3 AU, accreting pebbles composed of rock and ice. The rock falls to
the core, while the ice lodges in the envelope, subject to the
constraint that at each depth, the partial pressure of water ice
does not exceed the saturation vapor pressure. The goal is to determine
the critical  core mass of the planet (essentially the mass at which
substantial gas accretion starts to occur) for an envelope
enriched in heavy elements. The results show that this critical core mass
falls in the range 2--5 M$_\oplus$, lower than the values obtained
for envelopes of pure H/He, because of the higher mean molecular weight in
the envelope. This effect was previously predicted by \citet{ste84}.

In the present paper, two formation scenarios are considered for Kepler-36 c, one in which the planet
forms in situ at 0.128 AU, and  the other in  which it starts to form 
at an orbital distance of 1 AU, then migrates, during the later stages of
formation, 
to its present orbit. The arguments for and
against in situ models, as opposed to migration models, are summarized, 
along with relevant references, by \citet{bod14} and \citet{dan16}. 
The latter paper shows that the masses and radii of all the planets in the Kepler-11 
system, except Kepler-11 b,
can be matched (with standard core accretion) either by an in situ model or by a migration model.
After formation, 
our model planets are evolved, at constant heavy element mass, including
mass loss from  the H/He region of the envelope by XUV irradiation,
up to the stellar age. 
We  then compare models according to the standard core
accretion theory with those calculated with the updated version, both
for the in situ scenario and the migration scenario. With suitable choices for the initial
surface density of solids in the disk and for the efficiency of mass loss,  
reasonable agreement with Kepler-36 c's
observed mass and radius is found in all cases. For the case of Kepler-36 b, 
an updated formation calculation is performed in situ, leading eventually to complete
loss of the hydrogen-helium part of the envelope, in agreement with the works quoted above.

\section{Computational Method}

The calculations reported here use the following  prescription for
the deposition of heavy elements in the envelope. In all cases
the planet forms inside the ice line, so that the planetesimals
are composed of rock. As in \citet{pol96}, ablation and
breakup are included during a planetesimal's passage through the
envelope, and the amount of heavy elements deposited in each
mass layer at each time step is calculated. Breakup turns out to be
the main mechanism for mass deposition by accreting planetesimals.  The criterion
for breakup requires that the hydrodynamic (ram) pressure on the  incoming 
planetesimal must exceed its compressive strength, which is provided by self-gravity as
long as the radius of the object exceeds  a few tens of km \citep[e.g.,][]{dan15}].  
In practice, this criterion is met well above the surface of the solid/liquid core once
the mass exceeds a few M$_\oplus$. 
For reasons discussed
below, the heavy elements, now assumed to be vaporized, do not
mix to uniform composition,  but remain in the mass layer where they have been deposited.
Then, starting at the surface and working inwards, a calculation
determines, at a given layer,  whether the partial pressure of the rock vapor
($P_{\rm part}$)
exceeds the vapor pressure of rock at the surface temperature of  a planetesimal in
the layer. The vapor pressure (in dyne cm$^{-2}$)  is given by 
\begin{equation}
P_{\rm vap} = 3.92 \times 10^{13} \exp (-54700/T_s)~,
\label{eq:vapor}
\end{equation}
where $T_s$ is the temperature of the surface layers of a planetesimal \citep{dan15}. 
This expression is derived from data given in
\citet{mel07}.  
It   is approximate for SiO$_2$  and does not distinguish among the different 
phases of what is actually a polymineralic assemblage, plausibly dominated 
by olivine or pyroxene. We ignore the likely   presence of iron metal.  
Equation (\ref{eq:vapor}) does not distinguish whether the material is solid or liquid, 
but in practice the temperatures are such that liquid (or supercritical fluid) 
dominates the  SiO$_2$ accreted  after the envelope becomes sufficiently massive 
(even though the material arrives in the atmosphere 
as solid). The key features of our revised model are not sensitive to this 
choice of the vapor pressure curve, which could be wrong by an order of 
magnitude at some temperatures.

There is a wide range of estimates for the critical temperature ($T_{\rm crit}$)
for rock vapor; for a summary see \citet{mel07}. In our case it  is set to 5000 K; if $T > T_{\rm crit}$, $P_{\rm vap}$
is essentially infinite. This means that ``rock" and gas can mix in all
proportions above the critical temperature. 
If $P_{\rm part} > P_{\rm vap}$,  the excess heavy element material sinks to the
mass zone below, leaving  the considered layer saturated with rock
vapor. The calculation continues all the way to the solid/liquid core, which
can gain mass if the innermost zone satisfies $P_{\rm part} > P_{\rm vap}$. 
 The result, during the main solid accretion phase, can be the
structure of a ``wet adiabat", on which the partial pressure of the
heavy material is equal to the vapor pressure. 
Since, during this phase,  the gas accretion  rate is much less than the solid accretion rate (unlike the 
late-stage formation of giant planets), this prescription necessarily means that 
once the temperature reaches values for which the vapor pressure substantially 
exceeds the hydrogen pressure, the heavy element material that rains out differs 
little from the dense vapor immediately above -- they are both essentially ``rock".  
For example, a gas parcel that has a hydrogen/helium partial pressure of 1 bar ($10^6$  dyne cm$^{-2}$) 
at 5000 K will contain a rock partial pressure of $ 1.8 \times 10^8$ dyne cm$^{-2}$ according to Equation (\ref{eq:vapor}),
meaning the parcel is over 99\% rock by mole fraction (and over 99.9\% rock by mass). In
reality, the thermodynamic behavior near criticality must be two coexisting phases, 
one of which is droplets of molten rock containing dissolved gas, and the other of which is a fluid hydrogen 
phase containing large amounts of evaporated (fluid) rock. In practice, the amount of hydrogen that 
dissolves into the rock rain-out is small. Thus, this model is largely indistinguishable 
from standard models with respect to the way the elements are distributed.  The key  differences are: (1) 
the accreted rock is much hotter (eventually supercritical),  and (2)  heat may not readily escape. 
For clarity of presentation, we refer to the \emph{inner core} as the region that forms, during the
earliest accretion stages,  from silicate 
that arrives directly  as solid or liquid, and the \emph{outer core} as the almost pure silicate ``vapor" 
(actually supercritical fluid upon compression), formed by breakup of planetesimals, that overlays it.
In the following, we use $M_\mathrm{icore}$ and $R_\mathrm{icore}$ to refer to the mass and radius, respectively, 
of the \emph{inner} core. 
Just outside the outer core, 
there is a layer,  usually relatively thin, where the
rock mass fraction strongly decreases with increasing radius.
Above this region of non-uniform composition, the outer part
of the planetary envelope consists essentially of H/He, with
uniform solar composition.

The accretion rate of heavy elements (where $M_\mathrm{Z}$ is the total mass
in heavy elements) 
is given by the standard equation \citep{saf69}
\begin{equation}
\frac{dM_\mathrm{Z}}{dt} =  \dot M_\mathrm{Z}  = \pi R^2_{\rm capt} \sigma \Omega F_g~,
\label{eq:saf}
\end{equation}
where $R_{\rm capt}$ is the effective geometrical
capture radius for planetesimals, $\sigma$ is the mass per unit area of
solid material (planetesimals) in the disk, $\Omega$
is the planet's orbital frequency, and $F_g$ is the
gravitational enhancement factor to the geometrical capture cross section. 
The planetesimal
radius is taken to be 100 km, and $F_g$ is taken from \citet{gre92}.
The planetesimal accretion rate is very high in the inner region of a protoplanetary disk, and the
precise value of the planetesimal size  or the uncertainty in the
value of $F_g$ have  little effect on the outcome.
In practice $R_{\rm capt} > R_{\rm icore}$,
unless the envelope mass is 
negligible.  The presence of the gaseous envelope enhances the
capture radius, as determined by the procedures outlined in
\citet{pod88} and  \citet{pol96}. The value of $\sigma$ changes with time,
taking into account the  starting value for $M_\mathrm{core}$ as well  as the heavy-element mass 
subsequently deposited onto the planet, and
assuming that the feeding zone for solids includes the region
within 4 Hill radii ($R_H$) inside and outside the planet's orbital semimajor axis \citep{kar94}.

By the end of accretion, the inner (solid/liquid) core contains a relatively small fraction of
the total mass; most of the accreted heavy elements remain in the outer core, as vapor or
supercritical fluid with very small amounts of H/He. 
The inner core radius provides the inner boundary condition for the calculation
of the structure of the envelope (which includes the outer core). Structure models  
for the inner core are calculated
according to the procedure described in \citet{dan16}. The cores  are in
hydrostatic equilibrium, assuming an adiabatic interior [see Equations (30) and (31) in that paper],
 and are composed of pure silicates. Given the inner core mass,
and the temperature and pressure at the base of the envelope, the inner core
radius $R_{\rm icore}$ is provided in a lookup table, based on those models. The temperature and
pressure at the outer edge of the inner-core model match those at the base of the envelope. 

The structure of the envelope is calculated under the assumption of
hydrostatic equilibrium, spherical symmetry, and mass conservation. The
basic structure equations are given by \citet{kip90}. Added
mass of heavy elements is deposited locally, as described above, and
accreted H/He is added at the surface. If the planetesimals hit the inner core,  which
occurs only for a short time at the beginning of the calculation, 
the inner boundary condition for the luminosity is given by
\begin{equation} 
L_{\rm accretion} \approx \frac{GM_{\rm core} \dot M_\mathrm{Z} }{R_{\rm core}}~.
\end{equation}
Otherwise, the luminosity is zero at the inner boundary. In that case, the
mass and energy released by the accreted planetesimal are deposited at the breakup
point and smeared over two pressure scale heights.  The deposited energy in a
given zone is given by Equation (10) of \citet{pol96} and includes the
latent heat of vaporization. The energy equation 
includes this energy source term, heating, cooling, contraction, expansion,
and radiation from the surface. 

In regions where the composition is uniform, the Schwarzschild criterion
for convection is applied, and the adiabatic temperature gradient $\nabla_{\rm ad}$
is used. 
In the zones of the  envelope where the composition is non-uniform, the 
Ledoux condition for convection  is considered:
\begin{equation}
\frac{d \ln T}{d \ln P} > \left(\frac{d \ln T}{d \ln P}\right)_{\rm ad} 
-\frac{\chi_\mu~d \ln \mu}{\chi_T~d \ln P}~,
\label{eq:led}
\end{equation}
where $\mu$ is the mean molecular weight and 
\begin{equation}
\chi_\mu \equiv \left(\frac{\partial \ln P}{\partial \ln \mu}\right)_{\rho,T}
~~~~~~~{\rm and}~~~~~~~~\chi_T \equiv  \left(\frac{\partial \ln P}{\partial \ln T}\right)_{\rho, \mu}~.
\end{equation}
The structure of the layers of non-uniform composition is  found to be
stable against (ordinary) convection. In equilibrium models, the specific entropy increases
significantly outwards in such zones, as a result, in part, of the steep outward decrease in the
mean molecular weight (note  that the ``wet adiabat" does not have constant specific entropy). 
A further test was considered: take a point in a model where the ratio
of mass fractions of H/He and rock vapor is, say, 1:1. Given the density $\rho_1$ and pressure $P_1$ at
that point, adiabatically   decompress that layer to the pressure ($P_2$) of a higher layer where
the composition is all H/He (a finite displacement). The density $\rho_{\rm ad}$ after decompression is then
compared with $\rho_2$, the model density at $P_2$. If $\rho_{\rm ad} > \rho_2$, then the region is
stable against convection. All points that were tested in this manner, in the non-uniform
region, turned out to be stable. The actual temperature gradient then must be less steep than that
given by the left-hand side of Equation (\ref{eq:led}) but steeper than the adiabatic gradient, because
the layers are unstable according to the Schwarzschild criterion.  The actual value in such regions is
uncertain; in most of our calculations it is taken to be 90\% of the Ledoux condition. This condition is
commonly met, except in layers where the composition gradient is very steep and nearly discontinuous, in which
case the temperature gradient is set to less than the 90\% value to allow numerical convergence. Temperature
gradients in the non-uniform region can thus  be much steeper than the adiabatic.  Further, 
the energy transport in those layers is taken to be radiative, and no mixing of chemical
composition through those layers is considered. 

According to the evolutionary calculations of \citet{lec12}, during the formation phase, slow mixing processes,
such as double diffusive convection, are likely to involve long time scales compared with the
formation time  and are therefore neglected. We also neglect these  slow mixing
processes  during the cooling phase, although the much
longer time scales during that phase suggest that at least some compositional mixing may well occur, depending on the
parameters in the theory. The effect of these parameters on the degree of mixing should be examined in
future work. It is common practice (e.g., in modeling the atmospheres 
of giant planets) to think of the ``wet adiabat" as a convective state despite the compositional 
gradient. This state has a lower (i.e., less negative) temperature gradient than the dry adiabat 
because of the latent heat release that results in the upward adiabatic displacement of a saturated 
fluid element. In practice, this assumption of a convective state only makes sense if one thinks 
that there is perfect rain-out of condensate when a saturated parcel is lifted adiabatically.  
The conditions we encounter are enormously different from any of those considered in atmospheric 
dynamics because the compositional gradients are so large. It must be conceded 
that our understanding of these conditions is imperfect. There can be no doubt, however, that a 
supercritical mixture containing a compositional gradient cannot benefit from the latent heat 
release and rain-out, and its convective propensity is thus best assessed by the Ledoux criterion. 
Convective inhibition is further enhanced once the material is no longer an ideal gas, because 
the thermal effects on density are diminished then (i.e., $\alpha T <1$,  where $\alpha$  is the coefficient of thermal expansion).

The equation of state (EOS)   in the envelope is taken from tables of the equation of
state of SiO$_2$, mixed with various mass fractions of H/He, ranging from 0 to 1.
In the case of pure H/He, the tables reduce to the equation of state of
\citet{sau95}; the solar ratio of H to He is assumed. If there is a heavy element component, the tables are based
on the quotidian EOS of \citet{mor88}, as extended by \citet{vaz13}. The 
tables have been compared with the results from the SESAME EOS \citep{lyo92}
and the ANEOS \citep{tho72}, with good agreement. 

The Rosseland mean opacity during the formation phase 
includes the effects of dust grains, as calculated by \citet{dan13}
for the case of solar composition in the envelope. Tables
are provided as a function of temperature and density, taking into
account a number of grain species and a size distribution starting at
0.005 $\mu$m and ending at 1 mm. The number density $N_g$ for grains goes as
$N_g \propto r_g^{-3}$, where $r_g$ is the grain radius.  The grains are assumed
to be carried in to the envelope by the accreted nebular gas.
Once the grains evaporate, the gas opacities are taken from \citet{fer05} and 
\citet{igl96}. A diagram of the opacities, when grains are present, 
is shown in \citet{dan16}.  At temperatures below 2000 K the molecular opacities (with no grains) of
\citet{fre08} are added to the grain opacity.            They become
significant only in the final isolated phase, after accretion stops, when
the grains are assumed to settle into the interior and to evaporate.
In the inner region of the envelope, where the composition is 100\% rock vapor,
a  table is used with 100\% heavy elements, taken from data in the Opacity Project
archives \citep{sea94}. The temperatures in the
region where there is significant rock vapor are above 2000 K, and grains are
not considered. Below 3600 K, the molecular opacities of \citet{fre14} are
used with a ratio of metals to hydrogen of 100 [their Equations (3), (4), and (5)].
Between 3600 K and 3900 K, opacities are interpolated between the values of
\citet{fre14} and those from the Opacity Project table.  The high-metal opacities
are high enough so that the regions of the models with 100\% heavy elements
are fully convective; therefore the structure is insensitive to the opacity
values. In the transition region between 100\% heavy elements and solar composition,
which encompasses a small fraction of the mass,  opacities are interpolated
between the solar table and the high-Z table. The mass fraction of heavy elements is
determined for a given zone, and logarithms of the opacities from
these two tables  are interpolated linearly in the mass fraction.
A reduction in the assumed opacities, particularly at low temperature, 
 would increase the rate at which 
the  envelope could  cool      and therefore
increase the gas accretion rate. Tests of the sensitivity of the results to
the assumed opacities will be considered in future work.

The outer boundary conditions depend upon the phase of evolution. 
During the formation phase, nebular gas with solar composition is added to
maintain          the condition  that the planet outer  radius $R_p \approx 
R_\mathrm{eff}$, where 
$R_\mathrm{eff} = {\rm min}(R_B, 0.3 R_H)$
and  $R_B$ and $R_H$   are the Bondi radius and the Hill radius, 
respectively.
The constant  0.3 is consistent with  three-dimensional
numerical simulations of disk flow and accretion near an embedded 
planet  \citep{lis09,dan13}. 
 During the formation phase, the
temperature at $R_p$, $T_{\rm surf}$,  is  set to a constant 
value of 1000 K in the in situ scenario. In the migration
scenario, during the solid accretion phase at 1 AU, $T_{\rm surf} = 500$ K.
The density at $R_p$,  $\rho_\mathrm{neb}$, 
is determined from the assumed disk surface density:  $\rho_\mathrm{neb} = \sigma_g/(2H)$, where 
$\sigma_g$ is the gas surface density in the disk,  the scale height  $H=0.03~a_p$ and, initially,
$\sigma_g/\sigma_\mathrm{init} =200$               
($a_p$ is the distance of the planet from the star). 
 The density $\sigma_g$,  
in the cases of fixed $a_p$,  is assumed to decline
linearly with time up to 3.3 Myr, when disk accretion cuts off. 
In all of these simulations, the envelope masses, which by our definition include
the outer fluid core,  become significantly larger than the inner core
mass; nevertheless
the  phase of rapid gas accretion associated with the
growth of Jupiter-mass planets  never occurs. The important factor is the ratio of H/He
mass to total heavy-element  mass, which always remains small. During the isolation phase, photospheric
boundary conditions are applied, including the effects of irradiation from the central
star; details are given in \citet{dan16}, Equations (2) through (5). The equilibrium
temperature $T_\mathrm{eq}$ at the orbit of Kepler-36 c is taken to be 928 K (with an 
assumed albedo of 0.3).

A detailed calculation of migration of the planet, coupled with the evolution of
the protoplanetary disk, is beyond the scope of this paper but should be considered in future work.
Thus, a very simple model is employed.
Migration from 1 AU to 0.128 AU is assumed to take place on a characteristic time
scale of $1.5 \times 10^6$ yr.  This assumption is based on detailed calculations of
migration of models of the Kepler-11 system in \citet{dan16}. During the solid accretion
phase (Phase 1), the formation time is very short compared with the migration time.  Numerical
experiments on the initial assembly of the core, based on a standard core accretion model \citep{dan16}
at 1 AU, taking into account the structure and evolution of the disk,
show that by the time the core has accreted to 7 M$_\oplus$, its semimajor axis has decreased by
about    10\%. Thus, migration starts after the
completion of this phase, shortly after the onset of Phase 2 (during which slow accretion of both
gas and solids takes place), with
$M_p \approx 7$ M$_\oplus$ and an elapsed time of $\approx 10^5$ yr. 
During migration, the surface temperature   varies smoothly between 500 K  and
the ultimate $T_{\rm eq}$.
 The outer density $\rho_\mathrm{neb} \approx 4 \times 10^{-8}$ g cm$^{-3}$ at 1 AU, then increases
smoothly to $1 \times 10^{-6}$ g cm$^{-3}$ at 0.128 AU. During migration, gas accretion
continues to occur according to the usual condition $R_\mathrm{p} = R_\mathrm{eff}$. 
The quantity $R_\mathrm{eff}$ decreases as the planet moves inward because of the  decrease
in $R_H$, which determines the outer boundary condition during this phase. The heavy element accretion
rate is limited to a factor 2--3 less than the gas accretion rate, based on the results from
\citet{pol96} during Phase 2.
The decrease in $R_\mathrm{p}$ can lead to mass loss from the H/He envelope under certain
circumstances during this phase.

The isolation
mass for the heavy-element component of a non-migrating planet
is given by 
\begin{equation}
M_\mathrm{iso} = \frac{8}{\sqrt{3}}(\pi C)^{3/2} M_\star^{-1/2} 
\sigma_\mathrm{init}^{3/2} a_p^3~,
\label{eq:iso}
\end{equation}
where  
$M_\star$ is the mass of the central star, and  
$C \approx 4$, the number of  Hill-sphere radii defining the region, 
interior and exterior to the planetary orbit, 
from which the object  is able to capture planetesimals \citep{lis87}. 
Once $M_\mathrm{Z} \approx  M_\mathrm{iso}$,
the $dM_\mathrm{Z}/dt$ slows down drastically, but  gas accretion continues.
Thus, $\sigma_\mathrm{init}$ is chosen so that $M_\mathrm{iso} \approx M_p$,
the present mass of the planet, but note that after $M_\mathrm{iso}$ is reached
(which occurs before migration starts),
the planet's mass will increase with addition of gas and solids during 
Phase 2,             and will  decrease with gas mass loss, possibly during migration
and certainly during the isolation phase.  The calculations thus assume that
the accreted solids are present near the initial location of the growing planet; 
migration of solids from the outer
disk in to the formation location is not considered,
nor are possible changes in the accretion rate of solids caused by the planet's
own migration \citep{ali05,dan16}.

The rate of mass loss during the isolation phase, by irradiation of the planet by stellar X-ray
and EUV photons, assumes energy-limited escape \citep{wat81,erk07,mur09,lop12}
and is given by
\begin{equation}
\dot M_\mathrm{XUV}  \approx -\frac{\epsilon \pi R^3_\mathrm{XUV} F_\mathrm{XUV}}{K(\xi)GM_p}~,
\label{eq:ml}
\end{equation}
where $R_\mathrm{XUV} \approx 1.1 R_p$ is the radius at which most of the stellar
XUV flux is absorbed. The factor $K(\xi) = 1. -3/(2\xi) +1/(2 \xi^3) $ corrects for the
stellar tidal effect; $\xi = R_H/R_p$. The uncertain quantity $F_\mathrm{XUV}$ is taken
from \citet{rib05}. This flux is most intense for time $t <10^8$ yr and is given
by $F_\mathrm{XUV} = 3 \times 10^{-4}L_\star/(4 \pi a_p^2)$. After  that time
$F_\mathrm{XUV} = 3 \times 10^{-6} L_\star (5~{\rm Gyr}/t)^{1.23}/ (4 \pi a_p^2)$.
Here, $L_\star$ is the stellar bolometric luminosity, which varies with time according to
a theoretical stellar evolutionary track for    $M_\star = 1.07$ M$_\odot$. The track is
calculated with the program STELLAR \citep{bod07}; it starts in the pre-main-sequence
phase at $t=10^6$ yr and ends in the main-sequence phase at $t=7$ Gyr, where it matches,
within observational uncertainty, the present luminosity of Kepler-36. 
The generally assumed value of the efficiency
factor $\epsilon=0.1$, but other values, within about a factor 2, are considered.

At the time of disk dispersal, at the onset of the isolated phase, other mass-loss
mechanisms have been suggested \citep{owe16b,gin16}, driven basically by the loss
of surface pressure from the disk. The outer radius of the planet in those studies
is taken to be the Bondi radius; in our calculations for Kepler-36 c at disk dispersal, 
 the actual
radius, at 0.3 $R_H$, is a factor 10 smaller than $R_B$. The ``Parker wind"
mechanism \citep{owe16b} is not effective at such a radius; however this
possibility needs to be considered in detailed numerical simulations. For further
discussion, see \citet{dan16}, Section 2.3.

In summary, during the formation phase the following steps are taken during a time 
interval $\Delta t$: (1) calculation of mass and energy deposition by planetesimals, 
(2) calculation of rain-out and readjustment of mass and composition distributions, 
(3) solution of the full structure equations, given the updated composition
distribution, (4) in migration calculations, adjustment of the planet's semimajor axis,
and (5) addition  (or possible subtraction) of H/He at the surface. During the
isolation phase, at $a_p = 0.128$ AU, steps (2) and (3) are taken, and,
in addition, XUV-induced mass loss from the outer H/He layers is computed. A full
evolutionary sequence involves several thousand time steps $\Delta t$, of varying
length.

\section{Calculations and Results} \label{sec:results}

The calculations  start with an inner core mass of $ M_\mathrm{icore} \approx 0.5$
M$_\oplus$ and negligible envelope mass.
The  negligible envelope mass at the outset is consistent with this core
having formed quickly, since the associated accretion luminosity necessarily
leads to a high basal temperature for this envelope (thousands of degrees).
The ratio of the planets's outer radius (0.3 $R_H$ for an in situ calculation) to core radius is 
accordingly only about eight,  
implying only   about 3 orders of magnitude enhancement of the gas
pressure at the (inner)  core surface relative to the nebular pressure, insufficient to make
an envelope mass that is a significant fraction of an Earth mass. Therefore, 
$M_\mathrm{env} \approx  3 \times 10^{-3}$ M$_\oplus$. 
The remainder of the  formation phase is
calculated, with accretion of gas and solids (planetesimals), up through the lifetime
of the protoplanetary disk. Disk lifetimes are estimated to be a few Myr, with a range
from roughly 1 to 10 Myr \citep{ale14}. We arbitrarily take a value of  3.3  Myr.
The transition is then made to an isolated  (non-accreting) planet that  evolves
to the present state (7 Gyr) with evaporative
mass loss of the H/He envelope as a consequence of XUV irradiation \citep[e.g.,][]{mur09,lop12}.

The principal parameters are the surface density of solid material
in the disk  ($\sigma_\mathrm{init}$) at the time when the planet started to accrete,  and the
efficiency factor ($\epsilon$) in the formula for the XUV mass loss. There are numerous
other parameters involved in such simulations, including the equation
of state, the radiative opacity, the form of the surface boundary
condition, the treatment of zones with gradients in chemical
composition, the details of the calculation of migration, and others.
Here we do not do a systematic study of the effects of these parameters,
but use values consistent with previous work, except for the consequences of the
new physics (the possible dissolution of incoming planetesimals).
We  seek to establish the feasibility of explaining the properties of the planets with
model fits using the new physics. The surface density is adjusted to obtain an approximate fit
to the mass  of the planet at 7 Gyr, and then the
efficiency factor is fine-tuned to fit the radius, which also involves
a small adjustment in the mass.

For the case of Kepler-36 c, four model sequences are considered: $0.128(\mathrm{Rev})$, $0.128(\mathrm{Old})$, 
$1.00(\mathrm{Rev})$ and $1.00(\mathrm{Old})$.  The runs labelled (Rev) are calculated with
mass deposition in the envelope as described in the previous section. The runs
labelled (Old) assume, as in past calculations, e.g., \citet{dan16}, that planetesimal
material added to the envelope eventually sinks to the core, depositing mass and
energy at the core surface.  Otherwise, as far as possible, all other physical assumptions
and parameters are the same  in both types of runs. The  runs  labelled (0.128) assume that the
planet forms in situ at 0.128 AU from the star, while the  runs labelled (1.00)  start the planet
at 1 AU and migrate it to 0.128 AU while the protoplanetary disk is still present.
The starting time ($t_\mathrm{start}$) for all runs depends on the time $t_{0.5}$ to build a core of 0.5 M$_\oplus$,
as well as the time $t_\mathrm{pl}$ to form planetesimals of size 100 km. From Equation (\ref{eq:saf})
we estimate $t_{0.5} \approx 10^3$ yr at 1 AU, and it is even shorter at 0.128 AU.  
The time $t_\mathrm{pl}$ is unknown, but could well be longer
than $10^3$ yr; it depends on the detailed evolution of dust and gas in the disk. We arbitrarily set
$t_\mathrm{start} = 2 \times 10^3$ yr (for all runs); its precise value has practically no effect on the results and
conclusions of this paper. 
The cutoff time for accretion from the disk is about 3.3 Myr in all cases, and migration in 
the (1.00) runs starts shortly after the isolation mass has been reached, at $t \approx 10^5$ yr.

\begin{table*}
 \caption{Input Parameters and Results}\label{table:1}
 \centering
 \begin{tabular}{|l||cccc|}
 \hline\hline
 Run $\rightarrow$
& $0.128 (\mathrm{Rev})$ & $0.128 (\mathrm{Old})$
& $1.00 (\mathrm{Rev})$ & $1.00 (\mathrm{Old})$
   \\
 \hline
Disk solid $\sigma_\mathrm{init}$ (g cm$^{-2}$)
 & $1.18 \times 10^4$ & $1.085 \times 10^4$ & 196 & 190
  \\
 \hline
$\epsilon$ for $\dot M_\mathrm{XUV}$ & 0.08 & 0.22 & 0.04 &0.18 
   \\  
 \hline
 \hline
Final time (Gyr) 
 &    7.01  & 7.05 & 7.03   & 7.02                   
   \\
 \hline
Final $M_p$ (M$_\oplus$)           
 &    7.80  &   7.68  &  7.81   &  8.01                
   \\
 \hline
Final $M_\mathrm{icore}$ or $M_\mathrm{core}$  (M$_\oplus$) 
 &  1.30  & 7.00  &  1.87  &  7.32                        
    \\
 \hline
Final  env. $M_\mathrm{Z}$ (M$_\oplus$) 
 &   6.13   & --      & 5.65 &   --   
   \\
 \hline
Disk cutoff $M_\mathrm{XY}$  (M$_\oplus$)
 &   1.13   & 2.21  & 0.67  &  1.40                     
    \\
 \hline
Final $M_\mathrm{XY}$  (M$_\oplus$)
 &   0.37   & 0.68  &  0.29  &  0.69                     
    \\
 \hline
Final total $M_\mathrm{env}$  (M$_\oplus$)
 &    6.50  & 0.68  &  5.94  &   0.69                          
    \\
 \hline
Final $T_\mathrm{icb}$ or $T_\mathrm{cb}$  (K)
 &$1.75 \times 10^4$ &$2.20 \times 10^3$ & $1.54 \times 10^4$& $2.22 \times 10^3$
    \\
 \hline
Final $\rho_\mathrm{icb}$ or $\rho_\mathrm{cb}$ (g cm$^{-3}$)  
 &    8.00    &  0.46  &  7.59   &   0.46
   \\
 \hline
Final $\bar\rho_\mathrm{icore}$ or $\bar\rho_\mathrm{core}$ (g cm$^{-3}$)  
 &    8.55    &  6.59  &  8.35   &   6.38
   \\
 \hline
Final radius   (R$_\oplus$)                  
 &   3.66   &  3.74   &  3.72   &   3.72                           
   \\
 \hline
Final log ($L_\mathrm{int}$/L$_\odot$)
 & -10.85  &  -12.51  & -10.78  & -12.56
   \\
\hline
 \end{tabular}
\end{table*}

The parameters  and basic results for the runs are given in Table 
\ref{table:1}. The column headings 
in the table give the run identifiers. 
The first two  rows below the run identifiers  give the initial  
assumed surface density of solid material ($\sigma_\mathrm{init}$) in the disk, 
and the value of $\epsilon$ in Equation (\ref{eq:ml}). 
The initial gas surface density $\sigma_g$ in all cases is 200 times     
$\sigma_\mathrm{init}$. 
Note the very high values of $\sigma_\mathrm{init}$ that are required
to fit the mass of the present planet in the case of the  in situ  runs. 
The values are about 9 times higher than the corresponding surface density
\citep{chi13} in the typical minimum-mass extrasolar nebula (MMEN; their
Equation 4).
Note, however, that such a disk would still be gravitationally stable
\citep[see Figure 14 of][]{dan16}. In the case of the migration models,
the assumed values of $\sigma_\mathrm{init}$ are about 4 times higher than
the corresponding ones in the MMEN.

The bottom 12   rows  give results: the final values of time ($\approx 7$ Gyr), 
 final planet total mass $M_p$, the mass in the inner  core of heavy elements
$M_\mathrm{icore}$, for (Rev) models, along with the entire  core mass $M_\mathrm{core}$, for 
(Old) models,  the final mass of heavy elements $M_\mathrm{Z}$ in the        
envelope, the total mass in H/He at the time of disk cutoff (3.3 Myr), 
the final total mass ($M_\mathrm{XY}$) in H/He, the final total  
mass in the envelope $M_\mathrm{env}$, including both heavy elements and H/He, the final
temperature $T_\mathrm{icb}$, for (Rev) models,  at the inner core boundary, along with the corresponding
temperature $T_\mathrm{cb}$, for (Old) models, at the outer edge of the entire core, 
the final density $\rho_\mathrm{icb}$, for (Rev) models,  at the inner core 
boundary, along with the corresponding density $\rho_\mathrm{cb}$, for (Old) models, at the outer
edge of the entire core, 
the final mean density of the inner core ($\bar\rho_\mathrm{icore}$), for (Rev) models, along with
the final mean density of the entire core ($\bar\rho_\mathrm{core}$), for (Old) models,  
the final outer radius, and the final value of the intrinsic luminosity ($L_\mathrm{int}$).

\subsection{In situ model: Run 0.128(Rev)}

\begin{figure}[ht]
\centering%
\resizebox{\linewidth}{!}{\includegraphics[clip]{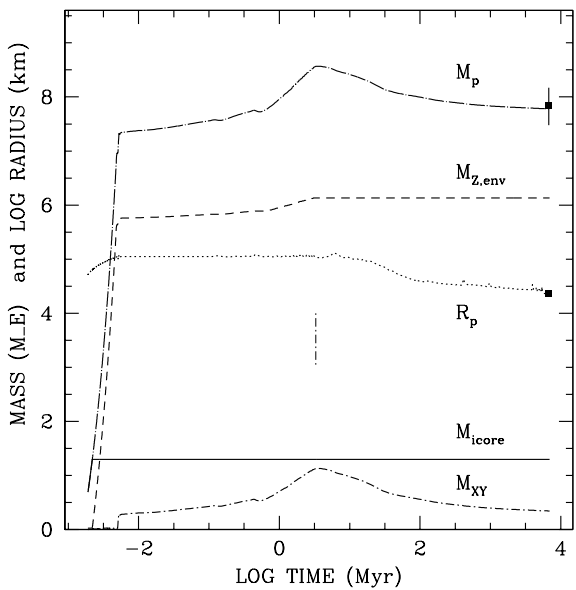}}
\caption{Evolution of Run $0.128(\mathrm{Rev})$. Upper (long dash-dot) curve: 
total mass $M_p$  (in M$_\oplus$);  dashed curve: total  
heavy-element mass in the envelope $M_\mathrm{Z,env}$; dotted curve: outer log radius $R_p$
(in km); solid curve: heavy-element inner core mass $M_\mathrm{icore}$; 
short dash-dot curve: hydrogen/helium mass in the envelope $M_\mathrm{XY}$; vertical dash-dot line:
time of disk accretion cutoff.
The observed mass of Kepler-36 c, with  error bars at 84\% confidence level, and the observed radius,
are given as filled squares. 
             }
\label{fig:1}
\end{figure}

\begin{figure}[ht]
\centering%
\resizebox{\linewidth}{!}{\includegraphics[clip]{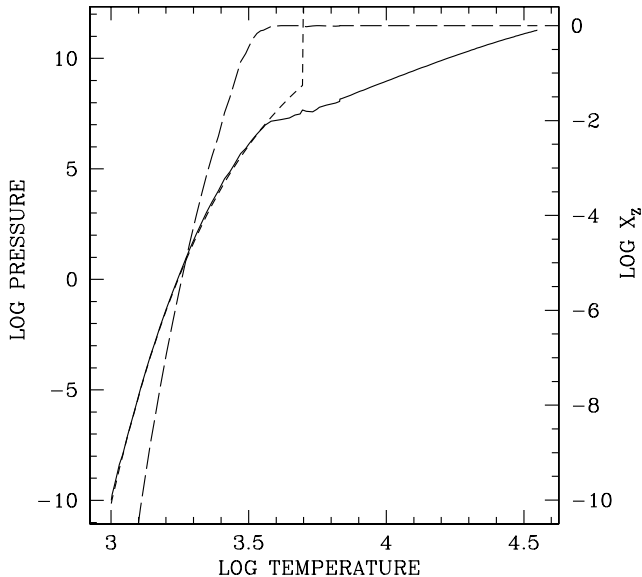}}
\caption{Structure of a model in Run $0.128(\mathrm{Rev})$ at a time (during the
runaway solids accretion epoch in Phase 1) when $M_\mathrm{icore} = 1.3$
M$_\oplus$,  
heavy-element mass in the envelope $M_\mathrm{Z,env}= 1.15$ M$_\oplus$, and
 hydrogen/helium mass in the envelope $M_\mathrm{XY} = 1.4 \times 10^{-2}$ M$_\oplus$.
Solid curve (left scale): partial pressure of the silicate vapor; short-dashed curve 
(left scale): vapor pressure
for the silicates; long-dashed curve (right scale): mass fraction ($X_Z$) of silicate vapor. Pressures are
given in dyne cm$^{-2}$, temperatures in K. Above the critical
temperature (uncertain but assumed  to be 5000 K) the vapor pressure is assumed to
become very high. 
             }
\label{fig:2}
\end{figure}

\begin{figure}[ht]
\centering%
\resizebox{\linewidth}{!}{\includegraphics[clip]{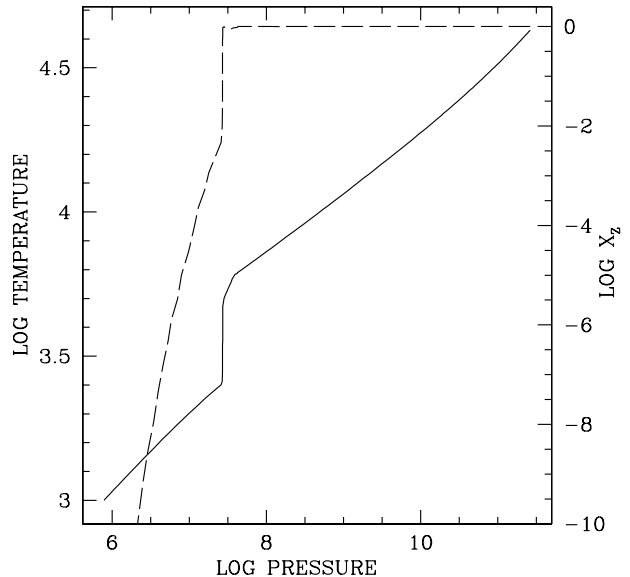}}
\caption{Structure of a model in Run $0.128(\mathrm{Rev})$ at a time when $M_\mathrm{icore} = 1.3$
M$_\oplus$,  
heavy-element mass in the envelope $M_\mathrm{Z,env}= 2.37$ M$_\oplus$, and
 hydrogen/helium mass in the envelope $M_\mathrm{XY} = 2.5 \times 10^{-2}$ M$_\oplus$.
Solid curve (left scale): temperature (K) as a function of total pressure (dyne cm$^{-2}$); dashed curve 
(right scale): 
mass fraction ($X_Z$) of silicate vapor. Note the very steep temperature and composition
gradients at log pressure = 7.5.}
\label{fig:2a}
\end{figure}

Masses and radius as a function of time for Run $0.128(\mathrm{Rev})$ are shown
in Figure \ref{fig:1}. The calculation starts with $M_\mathrm{icore} =0.40$ M$_\oplus$,
$M_\mathrm{env} =  2.2 \times 10^{-4}$ M$_\oplus$, with the envelope composed
entirely of H/He. 
In the preliminary phase of formation, the core accretes to 1.3 M$_\oplus$
in a time of only a few hundred years at the rapid solid accretion rate in the
inner disk. Up to  that point, a small amount of heavy elements lands in the envelope
through ablation, and some H/He is accreted, giving $M_\mathrm{Z,env} = 2.81 \times 10^{-2}$
M$_\oplus$ and $M_\mathrm{XY} = 9.85 \times 10^{-3}$ M$_\oplus$. 
Beyond that point, breakup of the planetesimals takes place in the envelope, 
 no further accretion onto the  inner core takes place, and all the
accreted planetesimals  remain in the envelope. The radiated luminosity during this phase
is $10^{-6}$ to $10^{-5}$ L$_\odot$, generally only 5--10\% of the rate of energy
deposition by planetesimals. The planetesimals release their energy interior to the layer
where the sharp molecular weight gradient occurs,  and because of the limited energy
transport across that layer, much of the deposited energy goes into heating and expansion
of the inner (high-Z) regions. During this solid accretion phase, the structure is fully
convective except in the layers with a composition gradient. The convective structure is
associated with the high nebular density ($\approx 2 \times 10^{-5}$ g cm$^{-3}$) and high
nebular temperature (1000 K) for the in situ case. An example of the structure during the
solid accretion phase is shown in Figure \ref{fig:2}. The partial pressure of the rock
vapor, the mass fraction of the rock vapor, and the vapor pressure are plotted as a 
function of temperature. An example of total pressure as a function of temperature during
this phase is shown
in Figure \ref{fig:2a}, emphasizing very steep composition and temperature gradients
in the layers where the mean molecular weight changes rapidly. In other regions, the 
gradient is adiabatic. 

 By the time of $5 \times 10^3$ yr,
all of the solid material in the feeding zone has been accreted, and Phase 2 starts.
At this time, the masses are: $M_\mathrm{icore} = 1.3$, $M_\mathrm{Z,env} = 5.64$, and $M_\mathrm{XY} = 0.027$,
all in Earth masses. The growth rate drops    drastically as the planet enters Phase 2. During this phase, 
the heavy-element  mass
increases  at roughly half the rate of  H/He  mass. 
In this connection, \citet{pol96} show that the accretion rate of solids ($\dot M_\mathrm{Z}$)
in this phase is related to the accretion rate of gas ($\dot M_\mathrm{XY}$)
by (their Equation 17) 
\begin{equation}
\dot M_\mathrm{Z} \approx {\left(2 +3 \frac{M_\mathrm{XY}}{M_\mathrm{Z}}\right)^{-1}} \dot M_\mathrm{XY}~.
\end{equation}
At the beginning of the phase, this expression gives a H/He accretion rate twice as fast as  
the heavy-element  accretion rate. At the end of the phase, when $M_\mathrm{XY}$ has increased to 1.13 M$_\oplus$,
the ratio is closer to 2.5, in reasonable agreement with the numerical results. 
\begin{figure}[hbt]
\begin{center}
\resizebox{\linewidth}{!}{\includegraphics[clip]{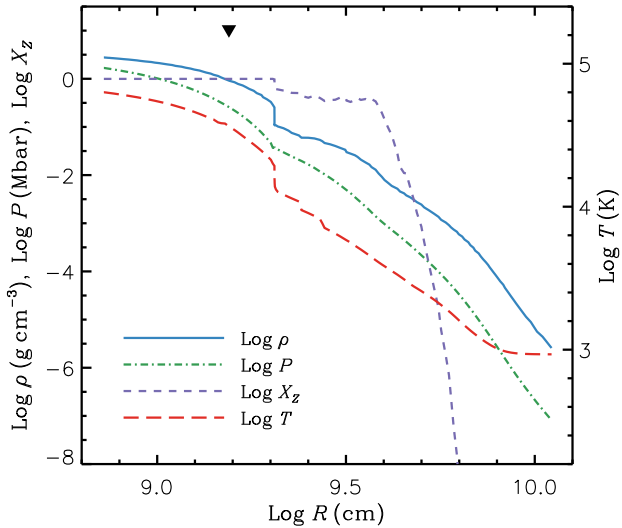}}
\caption{
Structure of the model in Run $0.128(\mathrm{Rev})$ at time 4.7 Myr,
soon after  the beginning of the isolated phase of evolution. 
Solid curve (left scale): log density in g cm$^{-3}$; 
dash-dot curve (left scale): log pressure in Mbar; 
long-dashed curve (right scale):
log temperature in K; short-dashed curve (left  scale): log $X_\mathrm{Z}$, the log 
of the mass fraction of heavy elements. Filled triangle: the half-mass point in
the envelope.  The mean density of the inner core is 5.62 g cm$^{-3}$. 
 The energy transport is mainly by convection; the
layers outside log $r=9.8$ are radiative.  The section of $X_\mathrm{Z}$ between 
log $r=9.3$ and log $r=9.6$  is a relic of Phase 2, during which the accretion rate
of H/He is 2 to 2.5 times greater than the solid accretion rate.  }
\label{fig:2b}
\end{center}
\end{figure}

At the beginning of Phase 2 there is
a brief readjustment, as the central regions, no longer supported by energy deposition from
planetesimals,  and still radiating at a rate controlled by the properties of the region of
non-uniform composition, contract significantly. The density $\rho_\mathrm{icb}$ (at the inner core boundary)
  increases from 0.6 to 3.6 
g cm$^{-3}$, and there is a brief burst in luminosity (to $\approx 10^{-3}$ L$_\odot$) as
the entire structure is forced to contract.  Thereafter, the luminosity declines
rapidly and remains at a typical value of $10^{-7.5}$ L$_\odot$ through Phase 2. 
As  a result of  the reduced luminosity, a radiative zone develops in the outer layers,
reaching inward to a temperature of 2000 K and to a radius about half the outermost
value, encompassing about 1\% of the total envelope mass (7\% of $M_\mathrm{XY}$).
 Disk cutoff occurs at $3.3 \times 10^6$
years with $M_\mathrm{Z,env} = 6.13$ and $M_\mathrm{XY} = 1.13$ M$_\oplus$, and with
$R_p = 17.7$ R$_\oplus$ ($1.12 \times 10^{10}$ cm). The temperature $T_\mathrm{icb}$  (just outside the
inner core boundary)  is $6.3\times 10^4$ K; the 
density (at the same point)  is 2.7 g cm$^{-3}$. The composition is uniform with 100\% Z out to a temperature
$T=2.08 \times 10^4$ K and radius $2.08 \times 10^9$ cm,   decreasing to 95\% at  $T=1.90 \times 10^4$ K at essentially
the same radius, to 50\% at $T=1.00 \times 10^4$ K at radius $2.44 \times 10^9$ cm, and to 1\% at $T=3000$ K, 
radius $4.62 \times 10^9$ cm. The structure of the model shortly after the cutoff is shown
in Figure \ref{fig:2b}. The outer radiative zone remains, extending inward to a temperature of
1500 K. 

\begin{figure}[htb]
\begin{center}
\resizebox{\linewidth}{!}{\includegraphics[clip]{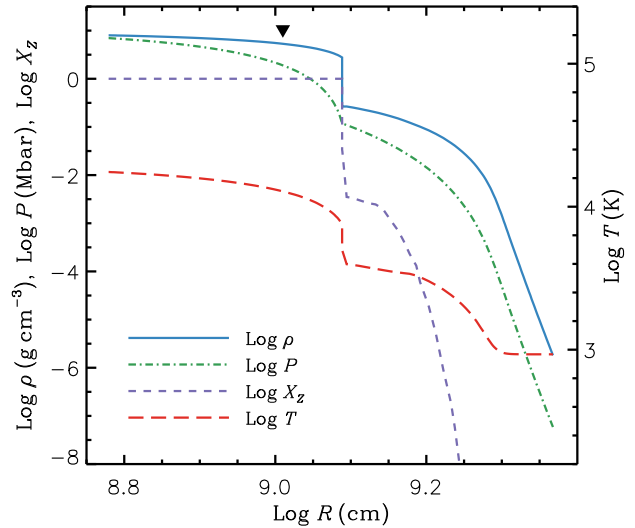}}
\caption{
Structure of the model in Run $0.128(\mathrm{Rev})$ at time $7.01  \times 10^9$ yr.
Symbols and curves as in Figure \ref{fig:2b}. The mean density of the inner core is
8.55 g cm$^{-3}$. }
\label{fig:22b}
\end{center}
\end{figure}

During the isolated phase, the parameter $\epsilon$ in Equation (\ref{eq:ml}) is set to
0.08. Initially, the high internal energy       and  average intrinsic luminosity around $10^{-9}$ L$_\odot$ 
combine to give a cooling time of $\approx 10^8$ yr. During the first $10^8$ yr, when   
the rate of mass loss is high, the radius decreases by a factor 2.6, and 0.56 M$_\oplus$ of
H/He is lost by photoevaporation ($\dot M_\mathrm{XUV} \approx 10^{-9}$ M$_\oplus$ yr$^{-1}$ at that time).
Later, the intrinsic luminosity declines to $\approx 10^{-11}$ L$_\odot$, the internal temperature cools by
a factor of $\approx 4$, 
the cooling time increases by an order of magnitude, and the rate of mass loss declines
significantly, by 2.5 orders of magnitude to $3 \times 10^{-12}$ M$_\oplus$ yr$^{-1}$ by the final time.

Between $t=10^8$ yr and $t=7 \times 10^9$ yr an additional 0.2 M$_\oplus$ is
lost. The final model planet, whose mass and radius agree quite well with that of the 
actual planet, has a total heavy element mass (including the inner core) of $M_\mathrm{Z}=7.43$
M$_\oplus$ and H/He mass $M_\mathrm{XY} = 0.37$ M$_\oplus$. The structure is still largely
convective, with an outer radiative zone including less than 1\% of the mass. The actual luminosity of the planet
is completely dominated by the re-radiation of stellar luminosity at the equilibrium temperature,
which gives log ($L/$L$_\odot$) decreasing from $-4.9$ to $-6.1$ as the planet contracts during the
isolation phase. This range holds for all cases discussed here.
The structure of the final model is shown in Figure \ref{fig:22b}.

\subsection{In situ model: Run 0.128(Old)}

\begin{figure}[htb]
\centering%
\resizebox{\linewidth}{!}{\includegraphics[clip]{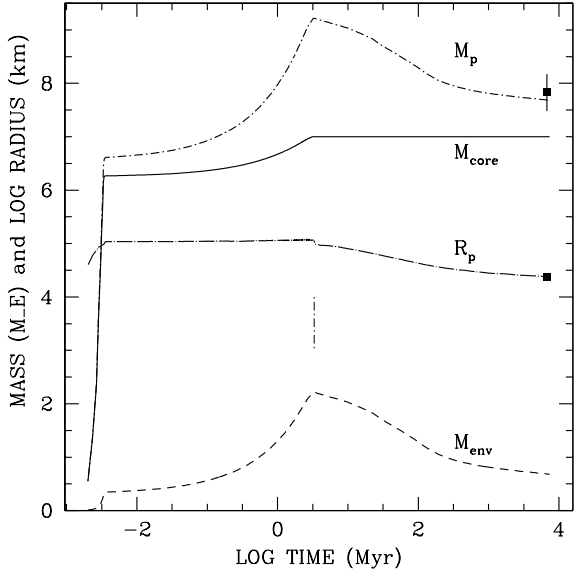}}
\caption{Evolution of Run $0.128(\mathrm{Old})$. Upper (short-dash dot) curve: total mass $M_p$  (in M$_\oplus$);  
solid  curve: core mass $M_\mathrm{core}$; long-dash dot curve: outer log radius $R_p$;
dashed curve: hydrogen/helium mass in the envelope $M_\mathrm{env}$; vertical dash-dot line: time of
disk accretion cutoff.
The observed mass of Kepler-36 c, with error bars at 84\% confidence level, and the observed radius,
are given as filled squares. 
             }
\label{fig:3}
\end{figure}

Run $0.128(\mathrm{Old})$ starts in situ at $\sigma_\mathrm{init}=1.085 \times 10^4$
g cm$^{-2}$, slightly lower than that in Run $0.128(\mathrm{Rev})$. 
The masses and radius as a function of time are given in Figure \ref{fig:3}.

At first, the total core mass ($M_\mathrm{core}$) increases very rapidly until it reaches 6.27 M$_\oplus$,
close to the isolation mass. At this point the $M_\mathrm{env} = 0.34$ M$_\oplus$. 
The temperature $T_\mathrm{cb}$  (at the base of the envelope)  is $1.03 \times 10^4$ K, much lower than the value of
$T_\mathrm{icb} = 5.8 \times 10^4$ K reached at a comparable evolutionary phase in Run 
$0.128(\mathrm{Rev})$. The structure is fully convective at this point. 

\begin{figure}[ht]
\centering%
\resizebox{\linewidth}{!}{\includegraphics[clip]{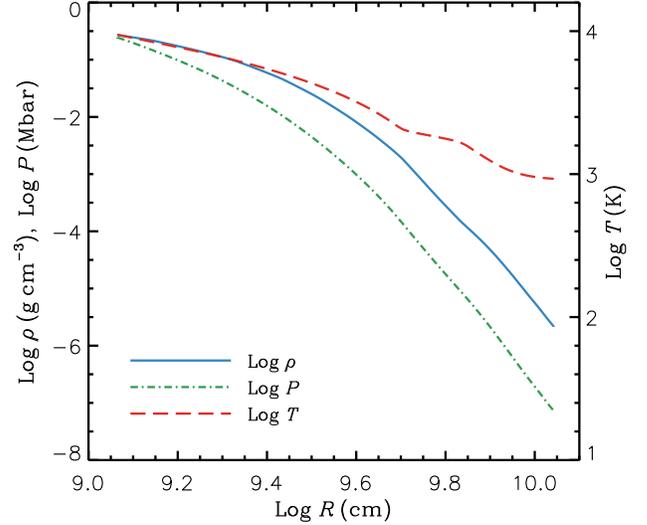}}
\caption{Structure of the H/He envelope in Run $0.128 (\mathrm{Old})$  at time 3.3 Myr (disk cutoff). 
Solid curve (left scale): log density as a function of radius;        
dash-dot curve (left scale):  log pressure as a function of radius;
dashed curve (right scale):  log temperature as a function of radius.
The mean density of the core is 6.42 g cm$^{-3}$. 
The structure is radiative outside log $r=9.7$; otherwise convective.              }
\label{fig:4}
\end{figure}

During the subsequent Phase 2, $M_\mathrm{core}$ increases
by 0.73 and $M_\mathrm{env}$ by 1.87 M$_\oplus$. The typical luminosity is
$10^{-7.5}$ L$_\odot$, about the same as in Run $0.128(\mathrm{Rev})$ during the
same phase. As in Run $0.128(\mathrm{Rev})$, a radiative zone develops in the outer
region, extending inward to $T=2000$ K. 
Disk cutoff occurs at time 3.3 Myr, with radius 17.5 R$_\oplus$, 
$T_\mathrm{cb} = 9430$ K, pressure at the core boundary $P_\mathrm{cb} =  0.245$ Mbar, 
and density $\rho_\mathrm{cb} = 0.271$ g cm$^{-3}$. 
All of these values are factors of a few lower than those in Run $0.128(\mathrm{Rev})$
 at the inner core boundary at disk cutoff. 
The core and envelope masses are, respectively, 7.0 and 2.21 M$_\oplus$.
The structure is plotted in Figure \ref{fig:4}. 

By the time of disk cutoff, this run was able to accrete twice as much H/He as was
possible for Run $0.128(\mathrm{Rev})$ at the same time.  The mean density
of the inner plus outer cores in Run $0.128(\mathrm{Rev})$, during the main phase of
gas accretion, is a factor 20 to 30 lower (with a correspondingly larger radius) than
the core density in Run $0.128 (\mathrm{Old})$.

At the beginning of the isolated
phase, the cooling time is $\approx 5 \times 10^7$ yr; the mass loss
efficiency factor is set to 0.22. At an age of $10^8$ yr, 
the temperature $T_\mathrm{cb}$ has decreased to $5.96 \times 10^3$ K and $M_\mathrm{env}$
to 1.28 M$_\oplus$, a loss  of 0.93 M$_\oplus$. An additional 0.60 M$_\oplus$
is lost up to the end of evolution at $7.05 \times 10^9$ yr. Near the
beginning of the isolated phase, the intrinsic luminosity, representing the cooling
of the planet,  is log ($L$/L$_\odot) = -8.5$, 
decreasing to log ($L$/L$_\odot) = -12.5$ at the final time.

\begin{figure}[ht]
\centering%
\resizebox{\linewidth}{!}{\includegraphics[clip]{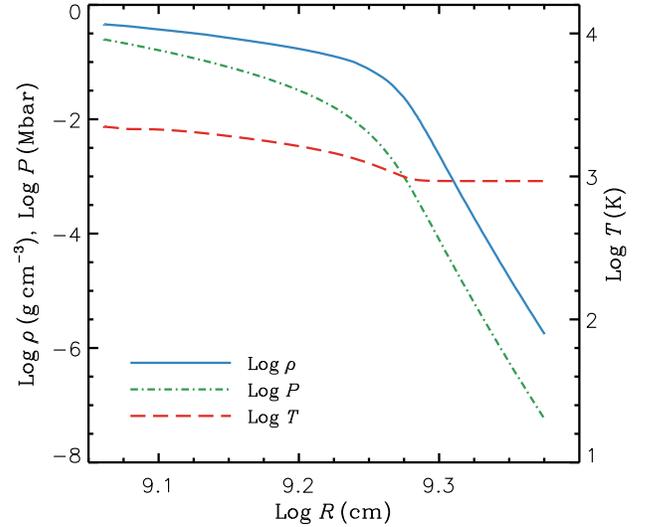}}
\caption{Structure of the H/He envelope in Run $0.128 (\mathrm{Old})$  at time $7.05  \times 10^9$ yr  (final model). 
Curves as in Figure \ref{fig:4}. The mean density of the core is 6.59 g cm$^{-3}$.}
\label{fig:44}
\end{figure}
The final model  (Figure \ref{fig:44}) has a radius of 3.74 R$_\oplus$, close to the upper limit of the
error bar for the planet \citep{dec12}.
The mass is 7.68 M$_\oplus$, in good agreement with that of the planet. 
The temperature $T_\mathrm{cb}$  has decreased to $2.23 \times 10^3$ K by this time, much cooler than
the value of $T_\mathrm{icb}=1.75 \times 10^4$ K at the end of Run $0.128(\mathrm{Rev})$. As a result of
the very low luminosity, the structure is fully radiative by this point.

In Run $0.128(\mathrm{Old})$
much more H/He  accretes into the envelope (2.21 M$_\oplus$) up to disk cutoff, as compared
with Run $0.128(\mathrm{Rev})$ (1.13 M$_\oplus$). The reason is that the outer core region
of the revised model is much hotter and less dense than are the corresponding mass elements in
the old model. 
Thus, in order to reduce $M_\mathrm{XY}$ to the point
where the radius agrees with that of the planet, a higher mass loss efficiency 
parameter, by over a factor of 2, is required.  Alternatively, we could have reduced the assumed 
lifetime of the disk and slightly increased $\sigma_\mathrm{init}$. Also, the old model, as a consequence
of its lower  total thermal energy, contracts faster than the revised one, reducing $\dot M_\mathrm{XUV}$
in comparison with the revised model.  To reduce the value of the required $\epsilon$, one must
change some other parameter, such as the ratio of the outer radius $R_p$ to $R_H$ (or the lifetime of
the gas in the protoplanetary disk). A run
was completed with $R_p/R_H = 0.25$ as compared with the normal value of 0.3. The main
effect is reduction of the accreted $M_\mathrm{XY}$ into the envelope. However, given the
same solid surface density, the total mass is reduced and there is a compensating
effect: the smaller radius results in slower mass loss during the isolation phase.
The end result was a model whose radius ($R_p=3.73$ R$_\oplus$) agrees well with that
of the planet, and whose   mass (7.49 M$_\oplus$) falls just within the error bar.
However, the efficiency factor,
adjusted to give the correct radius, 
has declined only slightly, from 0.22 to 0.18.

\subsection{Migration model: Run 1.00(Rev)}

\begin{figure}[]
\centering%
\resizebox{\linewidth}{!}{\includegraphics[clip]{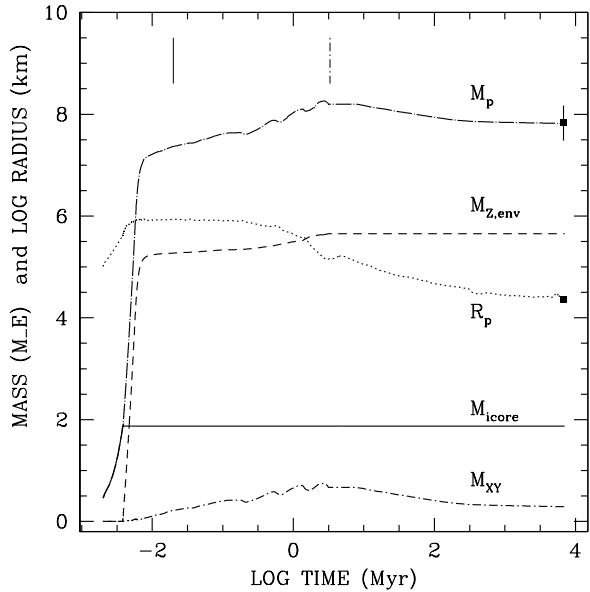}}
\caption{Evolution of Run $1.00 (\mathrm{Rev})$. 
Upper (long dash-dot) curve: total mass $M_p$  (in M$_\oplus$);  dashed curve: total  
heavy-element mass in the envelope $M_\mathrm{Z,env}$; dotted curve: outer log radius $R_p$
(in km); solid curve: heavy-element inner core mass $M_\mathrm{icore}$; 
short dash-dot curve: hydrogen/helium mass in the envelope $M_\mathrm{XY}$; thin  vertical solid line: 
time of onset of migration; thin vertical dash-dot line: time of disk accretion cutoff. 
The observed mass of Kepler-36 c, with  error bars at 84\% confidence level, and the observed radius,
are given as filled squares. 
             }
\label{fig:5}
\end{figure}

Masses and radius as a function of time for Run $1.00(\mathrm{Rev})$ are shown
in Figure \ref{fig:5}. The calculation starts with $M_\mathrm{icore} =0.46$ M$_\oplus$,
$M_\mathrm{env} =  5.4 \times 10^{-4}$ M$_\oplus$, with the envelope composed
almost entirely of H/He. 
At first, the core accretes to 1.81 M$_\oplus$
in a time of 1350 years, with a solid accretion rate $\approx 10^{-3}$ M$_\oplus$ yr$^{-1}$.
 At that point, $M_\mathrm{Z,env} = 1.5 \times 10^{-6}$  and $M_\mathrm{XY} = 3 \times 10^{-3}$
M$_\oplus$. Planetesimals continue to accrete onto the core until it reaches $M_\mathrm{icore} = 1.87$
M$_\oplus$.  
Beyond that point, breakup of the planetesimals takes place in the envelope, 
$M_\mathrm{icore}$ remains constant, and all the
accreted heavy elements remain in the envelope, forming the outer core.  During this phase, planetesimals are
deposited in the inner regions at radius $R_\mathrm{dep}$,  inside the layer where the steep composition
gradient occurs, at $R_\mathrm{dcont}$. For example, when the total envelope mass $M_\mathrm{env}= 1.5$ M$_\oplus$, 
$R_\mathrm{dep} = 5.29$ R$_\oplus$, $R_\mathrm{dcont} = 6.2$ R$_\oplus$, while  $R_\mathrm{icore} = 1.31$
 R$_\oplus$.  Also, when $M_\mathrm{env} = 2.67$ M$_\oplus$, 
$R_\mathrm{dep} = 7.13$ R$_\oplus$, $R_\mathrm{dcont} = 8.2$ R$_\oplus$, at  the same   $R_\mathrm{icore}$. 
 Only a fraction of the energy liberated at $R_\mathrm{dep}$ can be radiated
through $R_\mathrm{dcont}$,  and much of the deposited energy goes into heating and expansion of
the inner regions. For example, the luminosity radiated at the surface can be   as low as 1\% of 
 the rate of energy deposition in the interior.                    
This ratio varies with time. Interior and exterior to the layer with
the gradient, however,  the structure is convective. 

\begin{figure}[hbt]
\begin{center}
\resizebox{\linewidth}{!}{\includegraphics[clip]{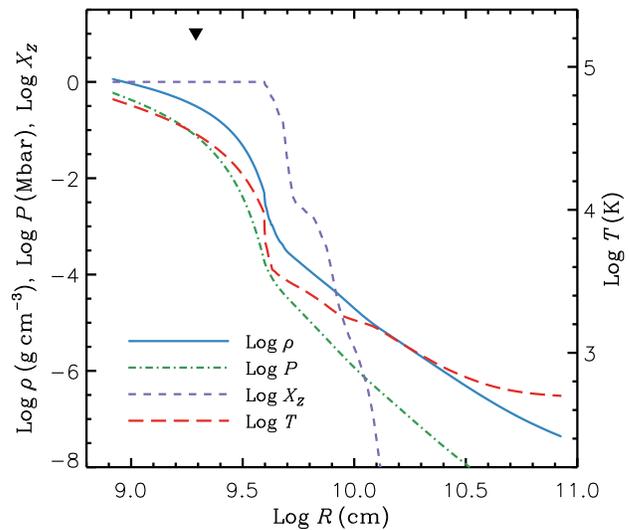}}
\caption{
Structure of the model in Run $1.00 (\mathrm{Rev})$ at time $1  \times 10^4$ yr,
at 1 AU just before the onset of migration.                  
Solid curve (left scale): log density in g cm$^{-3}$; 
dash-dot curve (left scale): log pressure in Mbar. The surface
value (not plotted) is $7.56 \times 10^{-4}$  bar. Long-dashed curve (right scale):
log temperature in K; short-dashed  curve (left  scale): log $X_\mathrm{Z}$, the log 
of the mass fraction of heavy elements. Filled triangle: the half-mass point in the
envelope. The mean density of the inner core is 4.63 g cm$^{-3}$. Note that the composition curve at
log $r=9.7$  is practically discontinuous; it is resolved by a few grid points. 
Interior to that point, the model is convective;  outside log $r=9.91$ it is radiative.
}
\label{fig:6}
\end{center}
\end{figure}

At the time of $1 \times 10^4$ yr,
practically all planetesimals  available  in the feeding zone have  been accreted, and Phase 2 starts.
At this time, the masses are: $M_\mathrm{icore} = 1.87$, $M_\mathrm{Z,env} = 5.26$,  $M_\mathrm{XY} = 0.14$, and
$M_p = 7.27$, all in Earth masses.  The outer radius is $R_p=133$~R$_\oplus$, as determined by $0.3~R_H$. 
The structure at this time  is shown in Figure \ref{fig:6}. As in previous cases, an outer
radiative zone develops. 

Shortly after  this time, migration starts. The rate of accretion of solids plays a much smaller role
in the overall energy budget during this phase, which is dominated by contraction and
accretion of H/He. As discussed in Section 2, the simple migration model neglects
the fact that the planet is migrating into a region that hasn't been mostly cleared of
planetesimals by the planet's own accretion (although the prior formation and migration of Kepler-36 b should have
done some clearing).
The growth time scale increases drastically to O($10^6$) yr, with heavy-element  mass
increasing at roughly half the rate of H/He mass.  At $t=0.8$ Myr, $a_p=0.6$ AU,
$R_p$ has decreased to 78 R$_\oplus$, $M_\mathrm{Z,env} = 5.46$ M$_\oplus$, and 
$M_\mathrm{XY} = 0.57$  M$_\oplus$. The H/He content reaches a maximum at $t=2.7 \times 10^6$ yr 
 when $a_p=0.17$ AU and $M_\mathrm{XY} = 0.75$ M$_\oplus$. Beyond that point, 
$M_\mathrm{XY}$ decreases as a result of  Roche-lobe overflow because of the 
 decreasing value of $R_H$. The luminosity during this phase declines gradually 
from $10^{-6.5}$ to $10^{-8.5}$ L$_\odot$ as a result of the decreasing accretion rate of
gas and solids as the value of the Hill radius decreases. Disk cutoff occurs at $3.3 \times 10^6$
years with $M_\mathrm{Z,env} = 5.65$ and $M_\mathrm{XY} = 0.67$ M$_\oplus$, and with
$R_p = 18$~R$_\oplus$ ($1.18 \times 10^{10}$ cm). The temperature $T_\mathrm{icb} = 6.36 \times 10^4$ K; the 
density $\rho_\mathrm{icb} =  1.47$ g cm$^{-3}$. The composition is uniform with 100\% silicates  out to a temperature
$T=2.1 \times 10^4$ K and radius $2.72 \times 10^9$ cm,   decreasing  to 50\% at $T=5.42 \times 10^3$ K and
radius $3.39 \times 10^9$ cm,  and to 1\% at $T=2960$ K and  radius $4.84 \times 10^9$ cm. 
\begin{figure}[hbt]
\begin{center}
\resizebox{\linewidth}{!}{\includegraphics[clip]{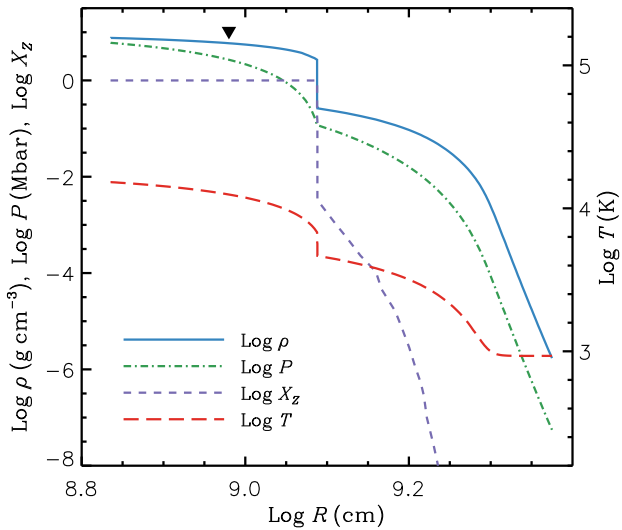}}
\caption{
Structure of the model in Run $1.00 (\mathrm{Rev})$ at time $7.03  \times 10^9$ yr.
Symbols and curves: as in Figure \ref{fig:6}. The mean density of the inner core is
8.35 g cm$^{-3}$. 
}
\label{fig:66}
\end{center}
\end{figure}

During the isolated phase, the parameter $\epsilon$ in Equation (\ref{eq:ml}) is set to
0.04.   During the first $10^8$ yr, an additional 0.25 M$_\oplus$ is lost, giving
$M_\mathrm{XY} = 0.42$ M$_\oplus$, $R_p=7.43$~R$_\oplus$, $L=2.0 \times 10^{-10}$ L$_\odot$, 
 and a cooling time of $8 \times 10^8$ yr. 
At this time,  $\dot M_\mathrm{XUV} \approx 8.8 \times 10^{-10}$ M$_\oplus$ yr$^{-1}$.
As in Run $0.128 (\mathrm{Rev})$, a radiative zone extends inward to $T=1500$ K. 
 Later, the luminosity declines to $\approx 10^{-11}$ L$_\odot$,
the cooling time increases by an order of magnitude, and the rate of mass loss declines
significantly, by a factor 400, to $2 \times 10^{-12}$ M$_\oplus$ yr$^{-1}$,      by the  final time.
 Between $t=10^8$ yr and $t=7 \times 10^9$ yr a further    0.13 M$_\oplus$ is
lost from the H/He envelope. The radius of the final model planet agrees quite well with that of the 
actual planet, as does the total mass. 
The  total heavy element mass (including the inner and outer cores) is $M_\mathrm{Z}=7.52$
M$_\oplus$,  and the H/He mass $M_\mathrm{XY} = 0.29$ M$_\oplus$.  The inner core of 1.87 M$_\oplus$ has
$R_\mathrm{icore} = 1.075$ R$_\oplus$ and mean density 8.35 g cm$^{-3}$. The region of almost
100\% heavy elements has radius 1.84~R$_\oplus$,  and  the  mean density  of the inner plus outer cores is
5.86 g cm$^{-3}$.

The mass of this final model is very close to     that of the in situ model
$0.128 (\mathrm{Rev})$. The temperature at the boundary between the inner and outer cores 
($T_\mathrm{icb}$) is similar ($1.54 \times 10^4$ K vs. $1.75 
\times 10^4$ K),  and the corresponding density is 
slightly lower (7.59  vs. 8.0 g cm$^{-3}$). These differences are presumably caused primarily by
the different masses of the inner cores. Both final models have outer radiative zones, extending
inward to $T=1850$ K  in the in situ case and to 1500 K in the present case; they include  less than 1\% of the envelope mass.
The structure of the final model for Run $1.00 (\mathrm{Rev})$ is plotted in Figure \ref{fig:66}.

\subsection{Migration model: Run 1.00(Old)} 

\begin{figure}[hbt]
\centering%
\resizebox{\linewidth}{!}{\includegraphics[clip]{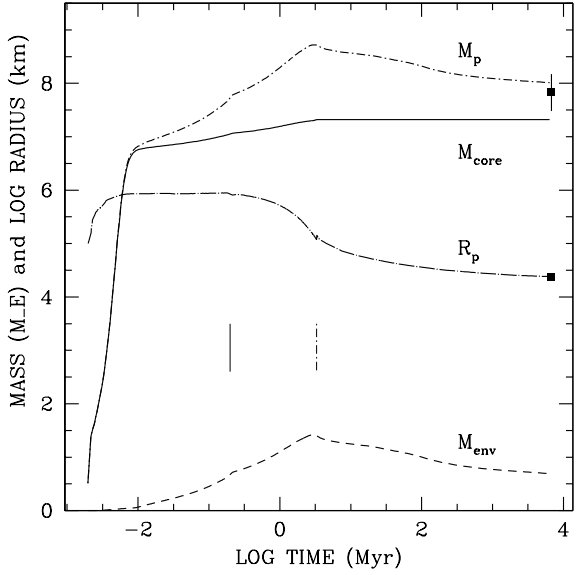}}
\caption{Evolution of Run $1.00(\mathrm{Old})$. Upper (short-dash dot) curve: total mass $M_p$  (in M$_\oplus$);  
solid  curve: total core mass $M_\mathrm{core}$; long-dash dot curve: outer log radius $R_p$;
dashed curve: hydrogen/helium mass in the envelope $M_\mathrm{env}$; thin vertical solid line: time of onset of
migration; thin vertical dash-dot line: time of disk accretion cutoff.
The observed mass of Kepler-36 c, with  error bars at 84\% confidence level, and the observed radius,
are given as filled squares. 
             }
\label{fig:7}
\end{figure}

Run $1.00(\mathrm{Old})$ starts at 1 AU with  $\sigma_\mathrm{init}=190$
g cm$^{-2}$, slightly lower than that in Run $1.00(\mathrm{Rev})$. 
Outer densities and temperatures are the same in the two runs.
The masses and radius as a function of time are given in Figure \ref{fig:7}.
At first, $M_\mathrm{core}$ increases very rapidly until, at $t \approx 7 \times 10^3$ yr, 
 it reaches 6.78 M$_\oplus$,
close to the isolation mass of 6.81 M$_\oplus$. At this point, $M_\mathrm{env} = 0.084$ M$_\oplus$. 
The temperature $T_\mathrm{cb}$ is $9.91 \times 10^3$ K, much lower than the value of
$T_\mathrm{icb} = 6.1 \times 10^4$ K reached at a comparable evolutionary phase in Run 
$1.00 (\mathrm{Rev})$. The lower mean molecular weight in the H/He envelope accounts for
much of this difference.  The luminosity during the solid accretion phase averages about $10^{-4}$ L$_\odot$,
corresponding to a mass accretion rate onto the core of 1 to 1.5 $\times 10^{-3}$ M$_\oplus$ yr$^{-1}$. 
Essentially all the energy deposited by the planetesimals is radiated away. The structure is
fully convective during this phase. 
\begin{figure}[hbt]
\begin{center}
\resizebox{\linewidth}{!}{\includegraphics[clip]{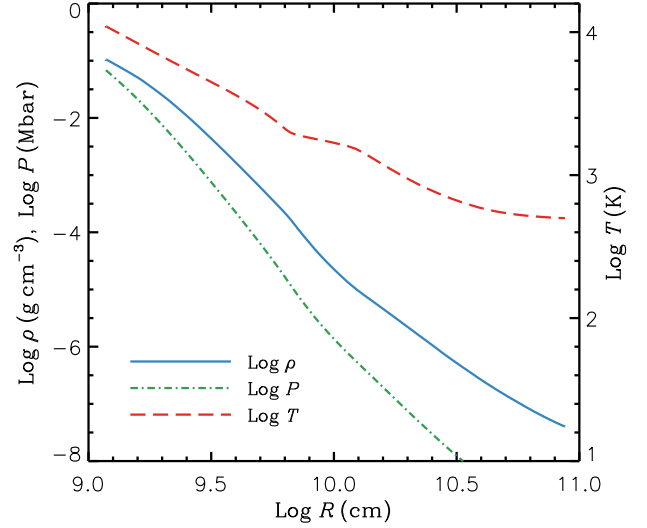}}
\caption{
Structure of the H/He envelope of the  model in Run $1.00(\mathrm{Old})$ at time $1.9  \times 10^5$ yr,
at 1 AU just before the onset of migration.                  
Solid curve (left scale): log density in g cm$^{-3}$; 
dash-dot curve (left scale): log pressure in Mbar. The surface
value (not plotted) is $7.18 \times 10^{-4}$  bar. Dashed curve (right scale):
log temperature in K. The mean density of the core is 6.13 g cm$^{-3}$. 
The change in slope of the temperature curve at log $r=9.8$
is the boundary between the inner convection zone and the outer radiative zone. 
}
\label{fig:8}
\end{center}
\end{figure}

Migration starts slightly later, with $M_\mathrm{core}=7.04$, $M_\mathrm{env}=0.63$ M$_\oplus$, 
and $R_p=137.6$ R$_\oplus$ (as determined by 0.3 $R_H$).
The structure of the model at this point is shown in Figure \ref{fig:8}; an outer radiative zone has
developed. 
At $t=0.8$ Myr, the planet has $a_p = 0.67$ AU  with  $M_\mathrm{core} = 7.17$ M$_\oplus$,
 $M_\mathrm{env}=1.02$  M$_\oplus$, and $R_p=92.2$ R$_\oplus$.  The luminosity during this phase
declines gradually from $10^{-6.5}$ to $10^{-8}$ L$_\odot$ as the accretion rate of gas and solids decreases.
More H/He mass is accumulated during migration than in the case $1.00 (\mathrm{Rev})$; however the mass loss
caused by Roche lobe overflow during the late stages of migration is negligible, only about
0.02 M$_\oplus$ [in Run $1.00 (\mathrm{Rev})$ it was 0.08 M$_\oplus$].
As is the case in the comparison between Run $0.128(\mathrm{Rev})$ and Run $0.128(\mathrm{Old})$, the
mass of H/He collected during the main gas accretion phase in Run $1.00(\mathrm{Old})$ is about twice
as great as that in Run $1.00(\mathrm{Rev})$, mainly because of the structure of the hot, low-density
outer core in the latter case.
Disk cutoff occurs at time 3.3 Myr, with the planet at its present orbital position and 
with radius 18.7 R$_\oplus$, 
temperature (at the core boundary)  $T_\mathrm{cb} =  1.04 \times 10^4$ K, pressure $P_\mathrm{cb} =  
0.16$ Mbar, and density $\rho_\mathrm{cb} =  0.20$ g cm$^{-3}$. 
The core and envelope masses are, respectively, 7.32 and 1.40 M$_\oplus$. During migration $M_\mathrm{core}$
and $M_\mathrm{env}$ have increased by 0.28 and 0.77 M$_\oplus$, respectively. The outer radiative zone
now covers  the outer 8\% of the mass. 
\begin{figure}[hbt]
\begin{center}
\resizebox{\linewidth}{!}{\includegraphics[clip]{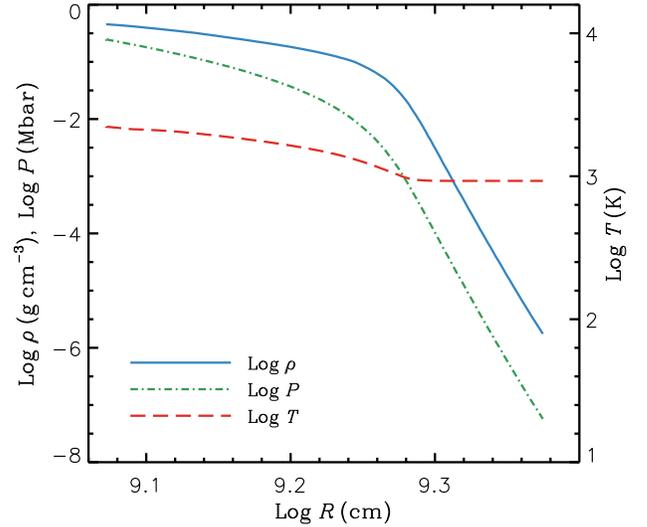}}
\caption{
Structure of the H/He envelope of the  model in Run $1.00(\mathrm{Old})$ at time $7.02 \times 10^9$ yr (final model). 
Curves as in Figure \ref{fig:8}. The mean density of the core is 6.38 g cm$^{-3}$.
}
\label{fig:88}
\end{center}
\end{figure}

\begin{figure}[hbt]
\begin{center}
\resizebox{\linewidth}{!}{\includegraphics[clip]{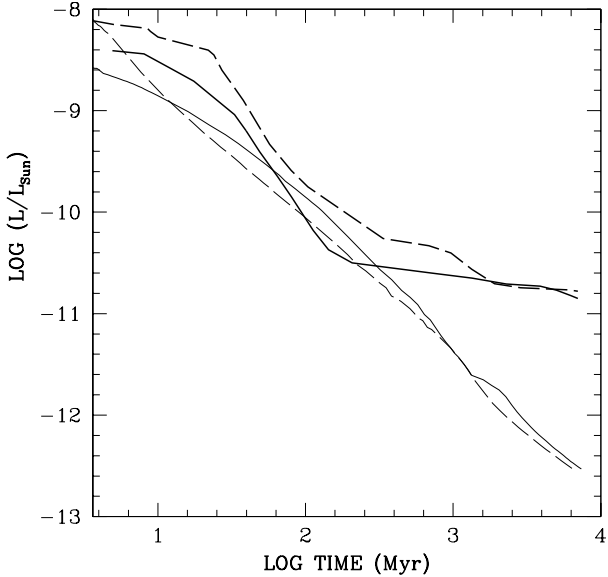}}
\caption{
Intrinsic luminosities as a function of time after completion of accretion. Thick solid line: Model 0.128(Rev).       
Thin solid  line: Model 0.128(Old). Thick dashed line: Model 1.00(Rev). Thin dashed line: Model 1.00(Old).
}
\label{fig:17}
\end{center}
\end{figure}

\begin{figure}[hbt]
\begin{center}
\resizebox{\linewidth}{!}{\includegraphics[clip]{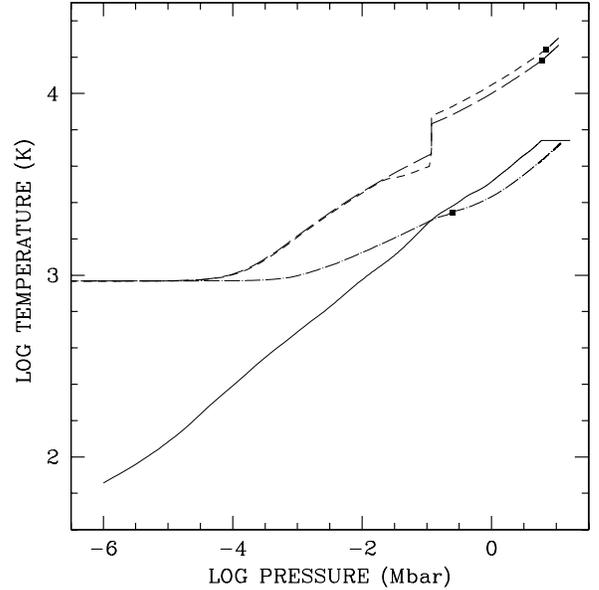}}
\caption{
Relation between pressure (in Mbar) and temperature (K) for Kepler-36 c models at time 7 Gyr, including the
inner core.  Thin solid line: 
model for the current Neptune from \citet{net13}. Note that the assumed composition is different from that
in this paper. Short-dashed curve: Run   0.128(Rev).  Long-dashed curve: Run 1.00(Rev). Long dash-dot curve: 
Run 1.00(Old). Solid squares: outer boundary of inner core, or the entire core in the (Old) model. Run
0.128(Old) is not plotted as the curve is practically the same as that for Run 1.00(Old).      
}
\label{fig:18}
\end{center}
\end{figure}

 At the beginning of the isolated
phase, the intrinsic luminosity is $10^{-8}$ L$_\odot$ and the cooling time is $\approx 1.0 \times 10^7$ yr. The mass loss
efficiency factor is set to 0.18. At an age of $10^8$ yr, 
the temperature  $T_\mathrm{cb}$ has decreased to $5.4 \times 10^3$ K and $M_\mathrm{env}$
to 1.01 M$_\oplus$, a loss  of 0.39 M$_\oplus$. The radius has decreased to 5.68 R$_\oplus$ and the 
luminosity to $10^{-10}$ L$_\odot$. The outer radiative zone has retreated, now covering only 1\% of the mass.
 An additional 0.32 M$_\oplus$
is lost up to the end of evolution at $7.02 \times 10^9$ yr. 
The final model  (Figure \ref{fig:88}) has a radius of 3.72 R$_\oplus$, in good agreement with that
of the planet. The mass $M_p= 8.01$ M$_\oplus$, is also in good agreement with that
 of the observed planet, with $M_\mathrm{core}= 7.32$ and $M_\mathrm{env}=0.69$ M$_\oplus$. The 
temperature $T_\mathrm{cb}$  has decreased to $2.22 \times 10^3$ K  and the intrinsic luminosity
to log ($L$/L$_\odot)  = -12.56$ by this time. About half of the envelope mass has been
lost through stellar XUV irradiation. The core radius is 1.85 R$_\oplus$ with mean density 6.38
g cm$^{-3}$. As in the case of Run $0.128 (\mathrm{Old})$, the structure is fully radiative, and envelope
masses, temperature $T_\mathrm{cb}$,  and density $\rho_\mathrm{cb}$  are essentially the same in the two cases.

The intrinsic luminosities as a function of time during the isolation phase for all four of the models
presented here are illustrated in Figure \ref{fig:17}. Note that the actual luminosities radiated by the planet
are many orders of magnitude higher. A comparison of the pressure-temperature relation in the structure of
three of the models is shown in Figure \ref{fig:18}. The inner core is included, whose structure is calculated
assuming an adiabatic temperature gradient. In the cases 0.128(Rev) and 1.00(Rev), the inner-core 
temperatures are likely above the melting curve of silicates \citep{mil15}, so the adiabatic assumption should be fully
consistent. In the case 1.00(Old), however, the temperatures are not high enough to satisfy that
condition, so the core may be semi-convective, at least in the outer shells. The core calculation for
1.00(Old) was rerun using (the convective-conductive)  Equation (29) of \citet{dan16} rather than (the adiabatic)
Equation (30).  The result is that the temperature at $r=0$ is considerably larger (about $10^4$ K versus
$5.29 \times 10^3$ K in the adiabatic case), but the pressure there is only slightly lower (by 0.05 Mbar). 
There is a negligible difference in the core radius.

\section{Kepler-36 b}
We now consider the question of why Kepler-36 b has such different
properties (e.g., much higher mean density) from Kepler-36 c, although its
orbit, at 0.115 AU, is not far inside that of Kepler-36 c. As mentioned above,
\citet{lop13} showed, on the basis of in situ  post-formation cooling
models, that Kepler-36 b could lose its entire H/He envelope as a result
of XUV irradiation from the star, while Kepler-36 c would not. The difference
is ascribed to the lower $M_\mathrm{Z}$ of b. Here we confirm that result by
providing a formation model for Kepler-36 b. It is assumed to form in situ
with an initial  core mass of 1.3~M$_\oplus$ and a nebular solid surface density
of $1.06 \times 10^4$ g cm$^{-2}$. Otherwise the assumptions and procedure
are the same as for Run $0.128(\mathrm{Rev})$. The orbital distance and surface 
density combine to give an isolation mass of 4.3 M$_\oplus$. The corresponding
values (Table 1) for Run 0.128($\mathrm{Rev}$) result in an isolation mass of
7.0 M$_\oplus$, leading to a significantly higher final mass for planet c. 

By the time of nebular cutoff at
3.3 Myr, the total mass is 4.48 M$_\oplus$, close to the actual
measured mass. The core mass is $M_\mathrm{icore}=1.3$~M$_\oplus$, the heavy-element  mass in the envelope is 3.05 M$_\oplus$,
and the H/He  mass is only 0.13 M$_\oplus$, 3\% of the total mass.
The quantity $\epsilon$ in the expression for XUV mass loss is set to 0.1 and $T_\mathrm{eq}$ to 978 K.
After a total time of $10^7$ years, $M_\mathrm{XY}$ has been reduced to
0.02 M$_\oplus$. The mass loss rate is $3 \times 10^{-9}$ M$_\oplus$ yr$^{-1}$, so
in another $10^7$ yr the entire H/He envelope would be lost. Note that the planet, at the beginning
of the isolated phase, has a higher thermal energy and a longer cooling time than would a model
planet calculated according to the standard (old) model. Thus, the revised model would have a larger
radius than the old during the early part of the cooling phase, and therefore would lose  H/He
mass more easily, given the same mass loss efficiency parameter. However, if $\epsilon$ is reduced,
the planet could possibly retain its H/He envelope. A calculation with $\epsilon=0.01$ shows that
the entire envelope would still be lost on a timescale of $2 \times 10^8$ yr. If it is further
reduced to 0.001, a low-mass H/He envelope ($\approx 0.1$ M$_\oplus$) is retained for over $10^{10}$ yr.
The borderline value of $\epsilon$, below which some H/He is retained for at least 7 Gyr, is
estimated to be 0.002.

After 7 Gyr of
cooling, the planet is expected, as observed, to be composed entirely of
heavy elements. Once sufficient H/He has been lost and the atmosphere is heavy-element
dominated, the energy-limited mass-loss expression (\ref{eq:ml}) is not applicable, as the rate-limiting
step for mass loss is the diffusion of H/He out of the atmosphere. 
 But, as the planet loses its envelope, the silicate
vapor in the deeper part of the envelope will supersaturate and rain out,
so eventually (nearly) all the H/He could be removed. This explanation applies only if
the photospheric temperature corresponds to a negligible silicate vapor
pressure.  Just before the standard calculation would predict that the H/He drops 
to zero, there will be a phase where diffusion-limited escape may apply, but the amount 
of gas left at that point is so small that it is not worth considering. If the deeper region is
uniform in composition (more precisely, if it has a homogeneous mantle
and a well-separated core), then it can cool very efficiently by
convection down to a state where it freezes at depth as well as at the
surface. The cooling time for this stage to reach something not that
different from the standard ``cold" picture is only a few hundred million
years. If the interior does have a composition gradient, however,
the planet could have
difficulty cooling and remain in an expanded state.

\begin{figure*}[t]
\centering%
\resizebox{0.6\linewidth}{!}{\includegraphics[clip]{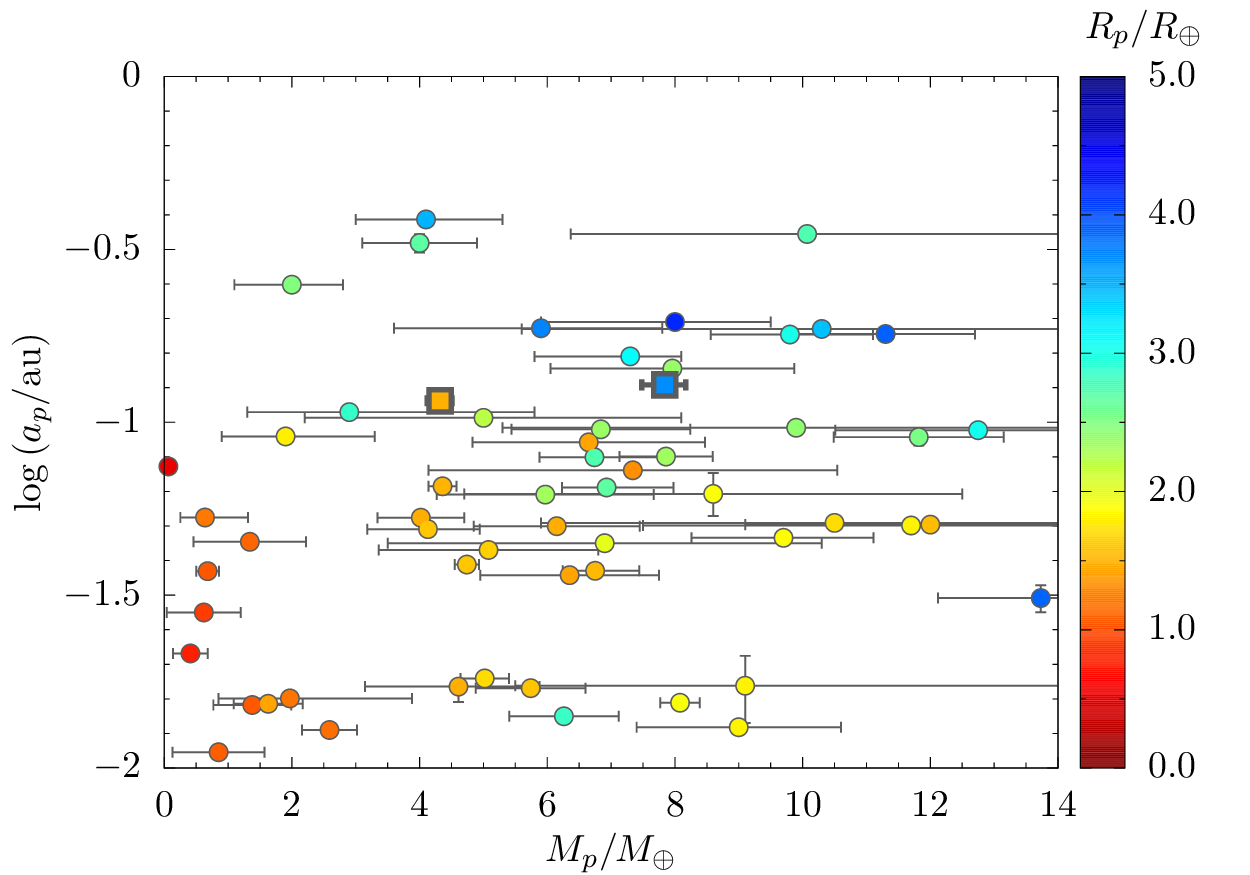}}
\caption{Radius and mass data from Figure \ref{fig:11} are shown   with the distance ($a_p$)  of planet
from star  plotted as a function of mass, with radius indicated by the color scale.
The two squares         indicate Kepler-36 b (less massive) and c.
}
\label{fig:12}
\end{figure*}

In this picture, the absence of H/He in Kepler-36 b is associated
with the fact that the planet's envelope, composed mainly of silicate vapor,
is unable to attract, during the formation phase, enough H/He to survive
the XUV irradiation during the isolated phase. \citet{owe16a} show that the absence of H/He in the
atmosphere of Kepler-36 b can be explained if, at the end of accretion, the heavy-element (core) mass was about 4.4 M$_\oplus$ and
the H/He envelope mass fraction was less than 10\%.  The formation calculations reported here are consistent
with their findings.
{As concluded by \citet{owe16a} and by \citet{lop13}, in the case of the Kepler-36 system, it is clearly the
difference in mass between the two planets, rather than the difference in location, 
that results in the much smaller radius for Kepler-36 b. The higher $M_Z$ of Kepler-36 c allows it to both
accrete and retain more H/He than did Kepler-36 b.
Figure \ref{fig:12} shows that
planets with small radii tend to be of   relatively low mass or located close to their star, or both. 

The comparison of the Kepler-36 b run and  Run 0.128($\mathrm{Rev}$) for Kepler-36 c shows that the differences
in mean density for the two planets can be explained. However, the calculations are based on the assumption
that both planets formed in situ at their current orbits. As discussed above, it is also    possible         
that Kepler-36 c formed at a larger distance and migrated inwards to its present orbit; Run 1.00($\mathrm{Rev}$)
also provides a fit to the present properties of the planet. If so, Kepler-36 b could have formed either in situ
or farther out in the disk, coupled with migration. In the former case, our calculations still show that the
differences between the two planets can be explained. In the latter case, the situation is more complicated,
because a detailed calculation has yet to be made. However, as an example, planet b could have formed at 0.75 AU
(interior to planet c) with an initial solid surface density of 250 g cm$^{-2}$ (higher than that for planet c),
giving an isolation nass of 4.3 M$_\oplus$, close to the measured mass. The shorter formation time during the
main solid accretion phase for planet b, along with the higher disk density,  would allow it to migrate inward ahead of planet c. Assuming that
the amount of H/He gas accreted by planet b up to disk dispersal was comparable to or up to a few times larger than that
in the in situ case, then it is still possible that the entire H/He envelope could have been lost by XUV
radiation during the isolated phase.

\section{Summary and Conclusions}

We investigate the formation and evolution, up to 7 Gyr,
of  (sub-Neptune) planets with total mass in the range 4--8 M$_\oplus$.
The models are compared with the observed properties of the 
planet Kepler-36 c, which orbits at 0.128 AU from a star of 
1.07 M$_\odot$, and planet Kepler-36 b, with an orbit at 0.115 AU. 
In the case of Kepler-36 c, we are able to adjust surface density and mass loss
efficiency so that models are found that agree quite well with both the mass
and radius of the planet at ages consistent with that of the star. 
 In the case of Kepler-36 b,  an in situ calculation shows that the
entire H/He envelope is lost,  with assumed surface density adjusted
to give the planet's observed mass, and with the mass  loss efficiency factor set
to the standard value of 0.1.
Our prescription assumes that the accreting planetesimals are
composed of rock, and we  take into account the breakup and
vaporization of the planetesimals as they interact with the
protoplanetary envelope. Dissolved rock vapor rains out to
lower levels if the partial pressure exceeds the local vapor
pressure. 
The main result is that the inner core (effectively pure silicate) 
of the planet remains at relatively low mass but is augmented by an 
outer core that is also almost pure silicate but arises from 
compressed silicate vapor that contains only small amounts of H/He
 and is much hotter than the same region of the planet in the older 
models. As a consequence, especially during and soon after the
accretion  stages, it is 
considerably less dense and causes the 
planet to have a somewhat   larger radius for the silicate-dominated 
portion alone. This silicate ``vapor" (actually a supercritical fluid) 
is concentrated in a region extending out to  as much as several inner core radii,
depending on the phase of evolution, 
and thus dominates the volume and mass of the total (inner plus outer) core.

The generally higher temperatures in the (Rev) models compared with the (Old) models
arise in part from the higher total envelope mass in the former case. In the (Old)
case  much of the accretional energy is radiated away, and the low-mass envelope
can store relatively little heat. In the (Rev) models, most of the mass lands in the
envelope, and the composition gradient results in limited heat loss by radiation,
so this envelope can store more of the accretion energy. Another effect arises from
the considerably higher mean molecular weight in the (Rev) case. To maintain
comparable pressures in the interiors of the two cases, as required for hydrostatic
equilibrium (actually the internal pressures in the (Rev) case are higher than
in the (Old) case, at equal total mass), higher temperatures
are required in the (Rev) case.

The outer core  is
bordered by layers in which the mass fraction of rock declines
sharply outwards; the composition gradient stabilizes the
layers against convection. Thus,  it is assumed that no chemical
mixing occurs between the inner rock-rich region and the
outer region, which is composed basically of H/He. Energy transport
through the region with the gradient is by radiation only. The
outer layers of H/He amount to only a small fraction of the
total mass, but a large fraction of the volume. 

The results are compared with models built according to the
old prescription, in which all accreted planetesimals end
up in the core, and the envelope has a uniform composition
of H/He. As in the old models, the revised models have a well-defined
core/envelope structure after 7 Gyr, but with different properties. 
Also considered for both old and revised models for Kepler-36 c,
are two different formation scenarios:  in the first, the
planet forms in situ at 0.128 AU; in the second, the initial
phase of rapid solid accretion occurs at 1 AU. Then, during
the subsequent phase of slow accretion of gas and solids, 
the planet migrates inward to its present orbital position. 
In all cases, during the isolated phase after disk dissipation,
mass loss from the H/He envelope is calculated, driven by
XUV irradiation from the central star. 
The main parameters that are varied to provide the fits are
the initial solid surface density in the disk at the formation
location, and the efficiency factor $\epsilon$  in the expression for
the XUV mass loss rate. The main conclusion is that our model, 
which accounts for dissolution of rocky planetesimal material
in the envelope of the forming planet,   accounts for the properties
of the planet Kepler-36 c, with suitable parameter choices for the initial solid
surface density in the disk and for the efficiency factor in the XUV mass-loss
formula,  with a 
lower-mass H/He envelope than required by the old models. 

A main feature of the revised calculation is the self-consistent
treatment of the composition distribution and the equilibrium
structure of the envelope of the planet, during its entire formation
and evolution. In order to concentrate on the effects of the
chemical composition and to allow the calculation of 
several   full formation/evolutionary sequences with a reasonable 
amount of computer time, a number of simplifications were made,
with respect to the state-of-the art simulations of planet
formation, e.g., \citet{dan14}.  For example, the additional
major refractory 
heavy-element component, iron, was not included. The dust opacity
relies on a fixed opacity table, rather than a detailed simulation
of dust settling and coagulation \citep{mov10}. The solid accretion
rate relies on a simple prescription, rather than the detailed
statistical treatment of the evolution and accretion of the
planetesimal swarm in \citet{dan14}. The temperature gradient in
the region of variable chemical composition is not well established
physically, and it is essentially parameterized. In the high-density
inner disk, it is possible that several planetary embryos can
form,  and later accrete to form one
object by giant impacts.  The impacts could modify the
formation process considerably and could cause mixing between
the silicate core and the outer H/He layers. Thus, the details of
the numerical results should be viewed with caution. Because of the
neglect of Fe, the comparisons with the observed properties of the
planet should be given less emphasis than the comparison between
the (Rev) models and the (Old) models.  The
general results of this paper could well stand up, subject to
more detailed simulations planned for the future.

The old and revised models, in both the
in situ case and in the migration case, form Kepler-36 c with 
conparable total amounts of heavy elements.
In the revised model, at the end of the calculation,  as a result of cooling
and contraction,    the heavy
elements are well-concentrated toward the center;  the size of that
region is only a few percent larger than the size of the core
in the old model, with lower mean density by a factor 1.12 [averaging
the (1.00) models and the (0.128) models]. The lower mean densities are
associated with  higher temperatures in the revised models.  
At earlier times, during the main gas accretion phase before $t=3.3$ Myr, 
the density in the silicate-rich outer cores in the revised models is only
3 to 5\% as large as in the  cores of the old models, the temperature at 
the base of the H/He-rich region is much higher, and
 the radius of the outer core is roughly a
factor 3 larger than the core radius of the old model at similar times. Thus, 
less H/He can be accreted in the revised models.
 The end result is that  models of Kepler-36 c according to the revised model have H/He 
envelopes of 0.29 and 0.37 M$_\oplus$, in models 1.00($\mathrm{Rev}$) and 0.128($\mathrm{Rev}$), 
respectively, only 4 to 5\% of the total
mass. In contrast, with the old model, the H/He mass
is about 0.7 M$_\oplus$, closer to 9\% of the total mass, as
also found by \citet{lop13} and \citet{owe16a}. 
At least two factors can account for this difference: (1) 
the higher temperature and 
lower density, during gas accretion,  in the
inner plus outer cores of the revised model compared to those in the
core of the old model; and (2) the  higher temperature and
lower density just outside the outer core in the revised model
compared with those just outside the core in the old model.
The transition zone, with the composition gradient,
plays a less important role, because the zone is relatively thin in
both mass and radius during the main gas accretion phase.     

It would be difficult observationally to distinguish
between the old and revised models, because both have significant
amounts of H/He at  the photosphere. Also,  their radiated luminosities
 at the present time would be very similar, completely dominated
by the stellar input and re-radiation.  We speculate that if a mechanism of slow
mixing of rock vapor occurs, during the long-term isolation phase, outwards
through the composition gradient into the largely convective H/He
layers, it might be possible to distinguish between the two models
on the basis of observed heavy-element ($Z$) abundances. 
The time scale
of, for example, double diffusive convection, is quite uncertain
\citep{lec12,mol17}, and this or related processes 
should be considered in future work. The complexities in the theory
are reviewed by \citet{gar18}. Nevertheless, the mixing of the rock vapor
outwards is much more likely during the long-term cooling phase than during
the formation (accretion) phase.
Note, however, that, first, the condensation of the refractories below
the observable photosphere must be taken into account,} and, second, the
enhancement of $Z$ abundances could also be caused by late accretion of planetesimals. 
The revised models presented
here may also change the speciation of oxygen and carbon in the observable
atmosphere, because the high temperature conditions that arise during
accretion change the speciation of these elements in the gas phase, as is
observed in the models of the deep atmosphere of Jupiter \citep{feg94}.       

Obtaining the fits to the observed mass and radius of the  planet turns out to be very sensitive
to the assumed parameters; fine tuning is required. For example,
in Run $1.00(\mathrm{Rev})$ the assumed value of $\epsilon$ was
0.04 (note that the generally assumed value is 0.1). The resulting
final planet radius $R_p$ was 3.72 R$_\oplus$ with $M_\mathrm{XY} =0.29$ M$_\oplus$. 
If $\epsilon$ was taken
to be 0.05,  $R_p= 3.19$ R$_\oplus$ and $M_\mathrm{XY} = 0.23$ M$_\oplus$. 
As another example, in Run $0.128(\mathrm{Rev})$
the solid surface density was $1.18 \times 10^4$ g cm$^{-2}$; 
the final mass was 7.80 M$_\oplus$ and the final radius 3.66~
R$_\oplus$. A run with the surface density $1.30 \times 10^4$
g cm$^{-2}$ gave, at 7 Gyr,  a mass of 10.7~M$_\oplus$ and a radius of
5.34 R$_\oplus$, both far too high to fit the planet. 
At the beginning of the isolation phase, this run achieved a total
mass of 12.6 M$_\oplus$, with $M_\mathrm{XY} = 3.26$ M$_\oplus$.
This model planet is somewhat short of the borderline, above which it 
would go  into
rapid gas accretion and  become a giant planet. 
The model fits found here are not necessarily unique; other
combinations of parameters could also match the observations. Such a parameter
study, which could involve numerous possibilities, is beyond the
scope of this paper. As examples, (1) if we allowed planetesimals to migrate relative to
the planet, then $\sigma_\mathrm{init}$ could be smaller, and (2)  if the nebula were to last longer, then
mass loss efficiency could  be higher.

Numerous discussions of the formation of hot Jupiters or super-Earth/sub-Neptune planets
in situ rather than ex situ have appeared in the literature. As summarized
by \citet{mor16}, the in situ scenario has two major problems.  
First,   the required solid surface density in the inner disk
is very high,  in our case  around 9 times higher than
that in the MMEN \citep{chi13}.   Second, 
in such a massive disk, the protoplanet is expected to migrate inwards, possibly
ending up in the star, or at least, inside the boundary of the magnetospheric
cavity, on a  time scale short compared with the disk lifetime. The first problem
could be solved to some extent if it is assumed 
that the  planet did form  in situ,  but  did not accrete from the local
disk mass, as was assumed here.  Rather, the planet  was built from protoplanetary cores \citep{war97},
or planetesimals \citep{han12}, or small rock particles (pebbles) \citep{tan16} 
 that migrated inward from the outer regions of the disk and collected at the current orbital
position of the planets. These processes would imply more gradual accretion of solids than we have
assumed here.  
In view of these problems, the possibility   that the planet formed at a larger distance should also
be considered. The actual formation location,
taken here to be 1 AU, is arbitrary but is consistent with our assumption that
the planetesimals are composed of rock. It is certainly possible that the
planet formed farther out, with an ice component. In that case, a much smaller
disk surface density would be sufficient to account for the planet's mass.
However, if Kepler-36 c formed beyond the ice condensation line, Kepler-36 b might 
well have also formed in that region, which would require an explanation of
how this rocky world lost all of its water in addition to its (much easier to lose) 
H/He.

\acknowledgments

Primary funding for the code development for this project was provided by the 
NASA Emerging Worlds program 15-EW15\_2-0007; the specific calculations for
the Kepler-36 planets were funded by 
the NASA Origins of 
Solar Systems Program grant  NNX14AG92G. We are indebted to
Ravit Helled, who provided the equation of state tables for compositions
that include silicates.  We thank
Stuart Weidenschilling for informative discussions, and Anthony Dobrovolskis
for useful comments on the manuscript.            
We thank the referee for a careful review which led to improvements in the
manuscript. 
Resources supporting the work presented herein were provided by the NASA 
High-End Computing (HEC) Program through the NASA Advanced Supercomputing (NAS) 
Division at Ames Research Center.


\end{document}